# DYNAMICS  INDUCED IN MICROPARTICLES BY ELECTROMAGNETIC FIELDS

***A conceptual introduction to nanoelectromechanical systems***

**Patricio Robles**

Escuela de Ingeniería Eléctrica, Pontificia Universidad Católica de Valparaíso, Chile

**Roberto Rojas**

Departamento de Física, Universidad Técnica Federico Santa María, Valparaíso, Chile

**Italo Chiarella**

Escuela de Ingeniería Eléctrica, Pontificia Universidad Católica de Valparaíso, Chile

# PREFACE

This is a review addressed to undergraduate students of electrical engineering and physics careers, and our aim is to present a pedagogical integration of some knowledge acquired in electromagnetism and electromechanical energy conversion courses, and to show how they can be used in actual technology applications, in particular for microscopic and nanoscopic systems.

In general, the contents of electromagnetic courses looks rather abstract to students at an intermediate level of the electrical engineering career and its applications are hard to be appreciated by them. The course of electromechanical energy conversion provides to the students a good link with processes of electrical engineering but in our opinion it is necessary a book containing a better integration and correlation of topics learned in these courses, such as energy and momentum transfer between electromagnetic fields and movile devices, at a macroscopic and at a microscopic level.

At a macroscopic scale, in general a good integration results in these courses for understanding how by means of variation in the energy stored in the magnetic field and interaction between rotating magnetic fields it is possible to produce mechanical or electric power for the operation of motors or generators, respectively. In general the programs of these courses give less emphasis on similar processes that are possible with the energy stored in the electric field as is the case, for example, of capacitors with movile plates or the interaction between electric polarized objects with rotating electric fields.

Furthermore it must be considered that technological development has made necessary to control the motion of elements with microscopic and nanoscopic dimensions, such as nanomotors and biological cells. One of the purposes of the present review is to explore this issue which in last years has originated a prolific research. In particular, it is worth to consider that during last year many research has been made in nanoelectromechanical systems (NEMS) which are a class of devices integrating electrical and mechanical functionality on the nanoscale.

In order to reach these objectives this review begins with an introduction where the motivation and main concepts presented in this review are summarized. In section 2 the concepts of polarizability and dielectric functions are presented. In section 3 the transfer of angular momentum from the electromagnetic field to isotropic and non-isotropic objects is analyzed, in particular the induced motion of an object with non-homogeneous polarizability. This is the case of a birrefringent element subject to a rotating electric field, where we study conditions for synchronous rotation. A similar analysis is made in section 4 for a magnetized nanowire in presence of a rotating magnetic field.

Section 5 discusses the analogy with a synchronous reluctance motor and section 6 presents some numerical results for illustrating the topics contained in this review.

An important part of this work is based on publications in collaboration of two of us (PR and RR) with Dr. Ronald Fuchs (R.I.P) and Dr. Francisco Claro, to whom we express our gratitude.

# TABLE OF CONTENTS

## 1. Introduction

In undergraduate courses dealing with electromagnetism and energy conversion, students acquire the concepts for understanding the interaction between electromagnetic fields ans how they can be applied to the analysis of some processes of energy transfer for producing motion by means of electromechanical devices such as transductors or electrical motors. For instance, it is explained that the variation of the energy stored in the magnetic field produced by the corresponding currents through a pair of coupled windings produces a torque tending towards the alignment of the corresponding axes. In the case of an induction motor, an undergraduate student learns that a continuous torque can be produced by the interaction between the rotating magnetic fields produced by the stator and rotor windings.

In this work it is shown that at the level of microscopic and nanoscopic dimensions, similar processes can take place. In fact, manipulation of nanoparticles, molecules and atoms by interaction with an electromagnetic field has been widely studied during the last few years, both from a theoretical as well as an experimental point of view [1-6], and several technological applications are being considered specially in biotechnology, for instance for controlling the motion of cells by using light [7]. A key factor for this optical manipulation is the transfer of linear and angular momentum from the light to particles immersed in a medium. In the representation of light as a beam of photons, each photon in a left/right-circularly polarized beam of electromagnetic waves carries $\pm 1$ units of angular momentum along the direction of propagation. In reference [8] the angular momentum of a photon in a circularly polarized beam is determined by evaluating the energy and angular momentum imparted by a classical plane wave to a point charge.

Optical and magnetic manipulations of particles can be obtained aplying both an external electric or magnetic field. For instance, in ref. [9] the motion of ferromagnetic nickel (Ni) nanowires under a rotating magnetic field is investigated both theoretically and experimentally and both synchronous as asynchronous rotations are analyzed. Ref.

[10] considers the motion of an active (self-propelling) particle with a permanent magnetic moment under the action of a rotating magnetic field, showing that below a critical frequency of the external field a synchronous rotation may be obtained.

In this work, we present a conceptual analysis of these subjects, starting from the generation of optical forces and torques over a system of particles. In particular we analyze the case of a non-isotropic particle immersed in a fluid and irradiated by circularly polarized light, determining the minimum intensity required for inducing a continuous rotation. For a nanoparticle initially at rest, we study the dynamics of the rotation and explain its analogy to the starting process of a synchronous reluctance motor. So, through an specific example we hope that undergraduate students may appreciate the close connection between concepts of two fundamental courses in electrical engineering careers, as Electromagnetism and Electromechanical Energy Conversion Theory. Furthermore, the example illustrates how to obtain a physical model suitable for the control of these rotation processes.

This work addressed to undergraduate students of electrical engineering and physics careers, is written in a pedagogical form so that each section contains an introduction remembering the key concepts involved. It is organized as follows: in section 2 the concept of polarizability of an object subjected to an electric field is presented. Section 3 explains how motion of particles can be induced by an electromagnetic field, forming clusters due to optical forces and torques produced by radiation. In particular in that section we present an analysis of the induced rotation of microscopic particles subject to a rotating electric field and we determine the conditions for synchronous rotation. Section 4 presents an analysis of the synchronous rotation of magnetic nanowires subject to an external rotating magnetic field and section 5 presents a comparison with the behavior of a synchronous reluctance motor. Section 6 includes some numerical examples.

## 2. Polarization and polarizability of a dielectric particle

When light passes through a material there is an electromagnetic interaction with the particles of the medium. This interaction is macroscopically manifested by two main effects: absorption of energy from the incident beam and scattering.

Considering a medium with atoms or molecules whose electrons acquire a motion due to the electric field associated to the incident electromagnetic wave, the absorption of energy may be understood using a phenomenological model of electric dipoles with negative charges whose positions oscillate with respect to the center of positive charges, with a frequency corresponding to that of the incident light. This oscillatory motion has a damping associated to the dielectric losses.

The scattering of light may be thought of as the redirection that takes place when an electromagnetic wave encounters an obstacle or non-homogeneity. The accelerated motion of the charges gives rise to radiation of electromagnetic energy in all directions producing secondary waves, process known as scattering.

A classical model for representing the optical response of a polarizable medium through which travels a monochromatic electromagnetic wave of frequency ω is the Lorentz model. In this model, due to the electric field vector of the wave some electrons bound to the nucleus of an atom perform harmonic oscillations. Therefore oscillating electric dipoles are formed and the corresponding electric dipolar moment is proportional to the electric field through the atomic polarizability $\alpha(\omega)$, with

$$\alpha(\omega) = \frac{e^2/m}{\omega_{oe}^2 - \omega^2 - i\Gamma_d}. \tag{1}$$

In this last equation $\omega_{oe}$ is the natural frequency of oscillation, $\Gamma_d$ is the absorption parameter and $m$ is the mass of the electron.

At a macroscopic scale the formation of electric dipoles in a dielectric material subjected to an applied electric field is described by means of the polarization vector $\vec{P}$ defined as the electric dipolar moment per unit of volume and related with the electric field and displacement vectors as

$$\vec{D} = \epsilon_0\vec{E} + \vec{P} = \epsilon_0\vec{E} + \chi\epsilon_0\vec{E} = \epsilon\epsilon_0\vec{E}. \quad (2)$$

In this last equation $\chi$ is the dielectric susceptibility and $\epsilon = 1 + \chi$ is the dielectric function which in general depends on the frequency $\omega$.

If $n$ is the volumetric density of molecular dipoles each one with Z electrons, the polarization vector is rewritten as

$$\vec{P} = \frac{nZe^2/m}{\omega_{oe}^2 - \omega^2 - i\Gamma_d}\vec{E}. \quad (3)$$

Therefore, we obtain the Lorentz model for the dielectric function:

$$\epsilon(\omega) = 1 + \frac{\omega_{pe}^2}{\omega_{oe}^2 - \omega^2 - i\Gamma_d} \quad (4)$$

where $\omega_{pe}$ is the plasma frequency given by $\omega_{pe}^2 = nZe^2/m\epsilon_0$. The derivation of eqs. (1) to (4) is shown in Appendix A.

From Eq. (4) it can be seen that a lossy dielectric and non magnetic medium has a complex refraction index $n_r = \sqrt{\epsilon(\omega)}$ whose imaginary part is associated to the attenuation of the intensity of an electromagnetic wave propagating in this medium due to absorption of energy.

For the magnetic permeability, see Ref. [11] where is used the following Drude-Lorentz model similar to that given by Eq. (4):

$$\mu(\omega) = 1 + \frac{\omega_{pm}^2}{\omega_{Tm}^2 - \omega^2 - i\Gamma_{dm}}, \quad (5)$$

where $\omega_{pm}$ is the magnetic coupling strength, $\omega_{Tm}$ is the transverse resonance frequency and $\Gamma_{dm}$ is the absorption parameter. The real part of $\epsilon(\omega)$ and $\mu(\omega)$ are negative for the following frequency ranges

$$\omega_{oe}^2 < \omega^2 < \omega_{oe}^2 + \omega_{pe}^2 \quad \text{for } \mathcal{R}e[\epsilon(\omega) < 0]$$
$$\omega_{Tm}^2 < \omega^2 < \omega_{Tm}^2 + \omega_{pm}^2 \quad \text{for } \mathcal{R}e[\mu(\omega) < 0]$$

When both $\epsilon(\omega)$ and $\mu(\omega)$ have negative real parts, the real part of the index of refraction is also negative and the medium has the behavior of a left-handed material (also known as a double-negative material) with the Poynting vector and the wave vector having opposite directions. For example, see ref. [12].

## 3. Analysis of a dielectric particle subjected to an electric field

The dynamics of nanoparticles, molecules and atoms interacting with an electric field has been widely studied both from a theoretical as from an experimental point of view [1-6].

The particles may be unpolarized initially but due to their mutual interaction and the interacting with an external field, they can acquire an induced electric dipole moment. As a consequence electric forces and alignment torques are produced, resulting in a movement of the particles, phenomena known like AC Electrokinetic which includes Dielectrophoresis and Electrorotation.

Dielectrophoresis effect corresponds to the movement of a particle subjected to a non uniform electric field [13, 14]. The Electrorotation effect corresponds to rotational motion of suspended dielectric particles due to the interaction with a rotating AC electric field [15]. In particular, in a previous work we have shown that an optical torque is induced by a circularly polarized beam, causing an asymmetric object to rotate uniformly if the light intensity exceeds a minimum value [16].

Understanding the dynamics of a system of particles due to the above phenomena is a very useful tool for several applications, for example the control of agglomeration or separation of proteins or living cells in suspensions.

One relevant issue is the determination of electric forces and torques over the particles. In reference [13] the time-averaged force between two polarizable nanoparticles subjected to a uniform oscillating electric field was obtained under the dipole approximation. This model, though delivering approximate results that are valid for interparticle separation not less than 1.5 diameters for spherical particles, gives a good phenomenological insight on the physics involved allowing a relatively easy interpretation of experimental results. In this approach, it was found the time averaged force between polarizable nanoparticles is non-conservative, non central, and at frequencies near resonance its magnitude may become greatly enhanced.

Because light carries linear momentum, the conservation laws require that when absorption of light occurs, the lost momentum is transferred to the irradiated object. Similarly, if circularly polarized light is absorbed, the angular momentum of such a field must cause an angular acceleration of the object. In this work we shall concentrate on the latter effect and discuss the rotation of small objects due to the torque caused by light. This issue has been analyzed in reference [16].

In this section we begin with the analysis of the induced force between particles that have acquired an electric dipole moment due to the action of an applied electric field, followed by the study of an isotropic particle which due to the transfer of angular momentum of a rotating electric field acquires a rotating motion. We continue with the case of a non-isotropic particle subject to a rotating electric field determining conditions for producing a synchonous rotation of this particle in a form similar to that of a synchronous reluctance motor. Up to here, all the analysis has considered only the dipolar approach, neglecting the effect of higher order induced multipoles. The effects of higher order multipoles is studied in subsection 3.4 and is complemented with appendixes B and C where mathematical detail are given.

## 3.1 Optical force between particles

An essential issue is the determination of the laser-induced electric forces between the particles. As stated before the time-averaged force in the dipole approximation was obtained in ref. [13], resulting that it is non-conservative, non- central, and at frequencies near resonance its magnitude is greatly enhanced.

In this subsection we study the motion of two identical polarizable nanospheres of mass *m* interacting in the presence of a laser field. We assume the particles are separated by a distance such that the dipolar approach is acceptable and higher order induced multipoles may be neglected. It is also assumed that the particles are neutral and the long wavelength limit applies so that the external field is of the form $\vec{E}_o e^{-i\omega t}$.

The spheres are immersed in a dielectric liquid, subjected to viscosity effects proportional to their velocities. It is assumed that forces due to the intensity gradient and Rayleigh scattering are small and may be neglected. Motion is referred to the center of mass (CM), and a *z*-axis is chosen along the electric field direction. Since the particles move on a plane we choose polar coordinates (r, θ) and (r, π+θ) for the position of the particles, as shown in Fig. 1. The radius of each sphere is *a*.

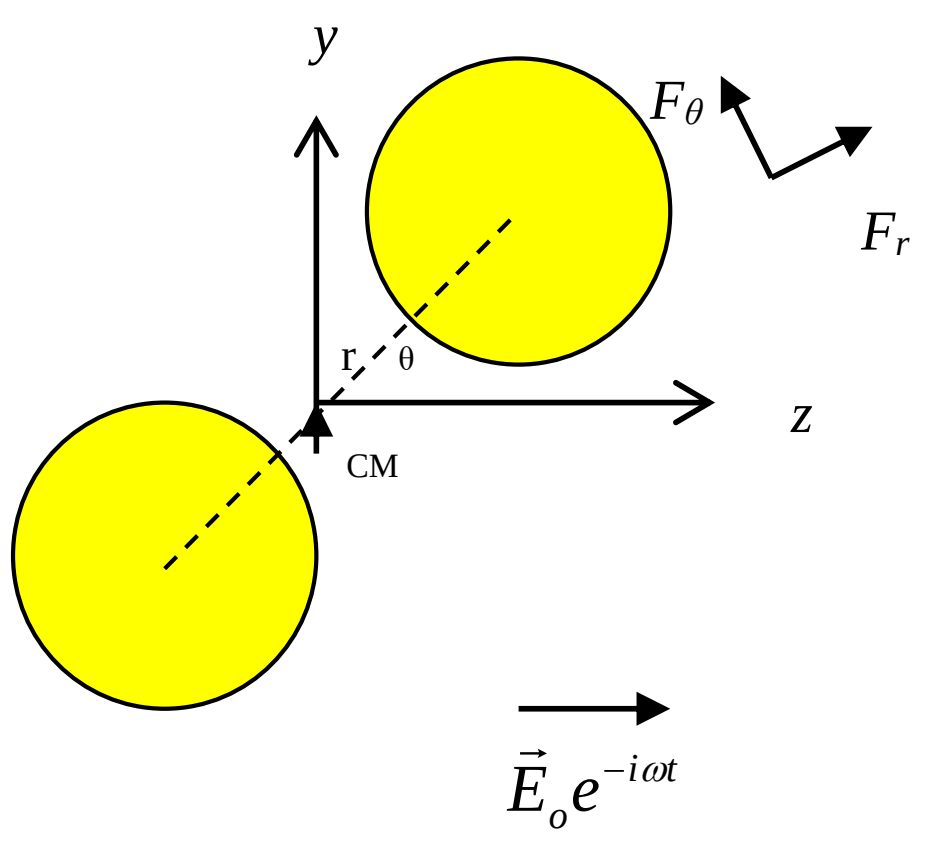

**Figure 1.** Two polarizable spheres excited by a fluctuating electric field along the z direction. References directions are indicated for the radial and angular components of the interaction force.

Due to the incident field the spheres become polarized and dipole moments are induced on them. As a consequence electric forces appear, with time-averaged radial and angular components given by: [13]

$$F_r(\vec{r}) = \frac{a^2 E_o^2}{96\sigma^4}\left[\frac{\sin^2\theta}{|n_R - u|^2} - \frac{2\cos^2\theta}{|n_A - u|^2}\right] \tag{6}$$

$$F_\theta(\vec{r}) = \frac{a^2 E_o^2}{96\sigma^4}\mathrm{Re}\left[\frac{\sin 2\theta}{(n_A - u)(n_R - u)^*}\right] \tag{7}$$

Appendix B includes a brief deduction of these equations and section 3.4 includes the description of a formalism for including higher order induced multipoles. Here $\sigma \equiv r/a$ is a dimensionless radial coordinate, such that 2σ represents the particle-particle separation. The material properties appear in the spectral variable $u = 1/[1-\varepsilon(\omega)]$ only, with $\varepsilon(\omega)$ the complex dielectric function of the spheres. The attractive and repulsive mode depolarization factors are given in the dipole approximation, respectively, by the same expressions given in [13]:

$$n_A = \frac{1}{3}\left(1 - \frac{1}{4\sigma^3}\right), \quad n_R = \frac{1}{3}\left(1 + \frac{1}{8\sigma^3}\right) \tag{8}$$

Notice that in the limit of large separation the radial force given by Eq. (6) is repulsive for $54.74^{\circ} < \theta < 125.26^{\circ}$. Furthermore, conditions of resonance can be deduced from the above equations. For example, if dielectric losses may be neglected so that the dielectric function is of the form $\epsilon(\omega) = 1 - (\omega_p/\omega)^2$ with $\omega_p$ being the plasma frequency, there will be resonance conditions for frequencies obtained from $u(\omega) = n_A$ and $u(\omega) = n_R$. More explicitely, for a pair of spheres of radii $a$ with centers separated by distance $r$, resonance frequencies are $\omega_{r1} = (\omega_p/\sqrt{3})\sqrt{1-(1/4\sigma^3)}$ and $\omega_{r2} = (\omega_p/\sqrt{3})\sqrt{1+(1/8\sigma^3)}$. For frequencies near $\omega_{r2}$ the repulsive effect predominates and the particles follow an unbounded trajectory.

We restrict the discussion of the light induced dynamics in this system to translational motion and use for its description Lagrange's equations, including a Rayleigh dissipation function [17]. In terms of the variables σ and θ one obtains the coupled equations for studying the dynamics of this system with $u'$ and $u''$ the real and imaginary parts, respectively, of the spectral variable $u$:

$$\frac{1}{\sigma^4}\left[\frac{\sin^2\theta}{(n_R-u')^2+u''^2}-\frac{2\cos^2\theta}{(n_A-u')^2+u''^2}\right]+F_0=\frac{d^2\sigma}{d\tau^2}-\sigma\left(\frac{d\theta}{d\tau}\right)^2+\lambda_r\frac{d\sigma}{d\tau} \qquad (9)$$

$$-\frac{1}{\sigma^4}\frac{\sin 2\theta\ \cos(\alpha_2-\alpha_1)}{\left[(n_A-u')^2+(u'')^2\right]^{1/2}\left[(n_R-u')^2+(u'')^2\right]^{1/2}}=\sigma\frac{d^2\theta}{d\tau^2}+2\frac{d\sigma}{d\tau}\frac{d\theta}{d\tau}+\lambda_\theta\sigma\frac{d\theta}{d\tau}. \qquad (10)$$

We have added an intrinsic radial force component $F_o$, that may be due to a Van der Waals interaction or to other effects. We assume the long range value of such force to be negligible in the presence of a strong laser field however, and take only a contact interaction of the form $F_0 = A\theta(\sigma-1)$, where $\theta(x)$ is the step function, and the amplitude $A$ is given the value 100 for convenience. The time scale is $T=\sqrt{96ma/a^2E_o^2}$ so that in such units the time is denoted by $\tau = t/T$. Damping is characterized by the parameters $\lambda_r = k_r\tau/m$ and $\lambda_\theta = k_\theta\tau/m$, with $k_r = k_\theta = 6\pi a\eta$, and $\eta$ being the viscosity coefficient of the medium surrounding the spheres. Finally, $\alpha_1$ and $\alpha_2$ are the polar angles of the complex quantities $n_A-u$ and $n_R-u$.

An example of light induced dynamics for this pair of particles is shown in Fig. 2. It was extracted from ref. [7] and correspond to a system of two spheres of gold, with radius 10 nm and an applied electric field of amplitude of 10 kV/m. The dielectric function of the sphere is given by $\varepsilon(\omega)=\varepsilon_b-\frac{\omega_p{}^2}{\omega^2+i\gamma\omega}$ with $\varepsilon_b = 9.9$, $\hbar\omega_p$ and $\hbar\gamma$ with values of 8.19 $eV$ and 0.027 $eV$ (expressed in equivalent units of energy of a photon, with $\hbar = h/2\pi$ being the Plank's constant). The particles are released from rest and followed in their motion on the y-z plane, resulting in trajectories as those shown in Fig. 2. The coordinate system is shown fixed at the center of one of the spheres. Dots along the path followed by the second particle are at equal time intervals $\Delta t$ of values included in the figure. The laser field wavelength is 1050 nm and the viscosity, null. Shaded (unshaded) regions correspond to initial positions leading to unbounded (clustering) trajectories, with a boundary placed at about $\theta_{max} = 63^o$. For determining this boundary a trajectory was considered unbounded if after a time greater than $T$=5000 (in units of $\tau$) the radial velocity was still positive, meaning a diverging path. For comparison, the slightly smaller critical angle for attraction $\theta_c$=54.74$^o$ is also shown in the figure, by means of a dashed straight line. As illustration, two trajectories are included, one bounded (full line, initial point at $\sigma_o$ = 2.5, $\theta_o$ = 45$^\circ$) and one unbounded (dashed line, initial point at $\sigma_o$ = 2.5, $\theta_o$ = 75$^\circ$).

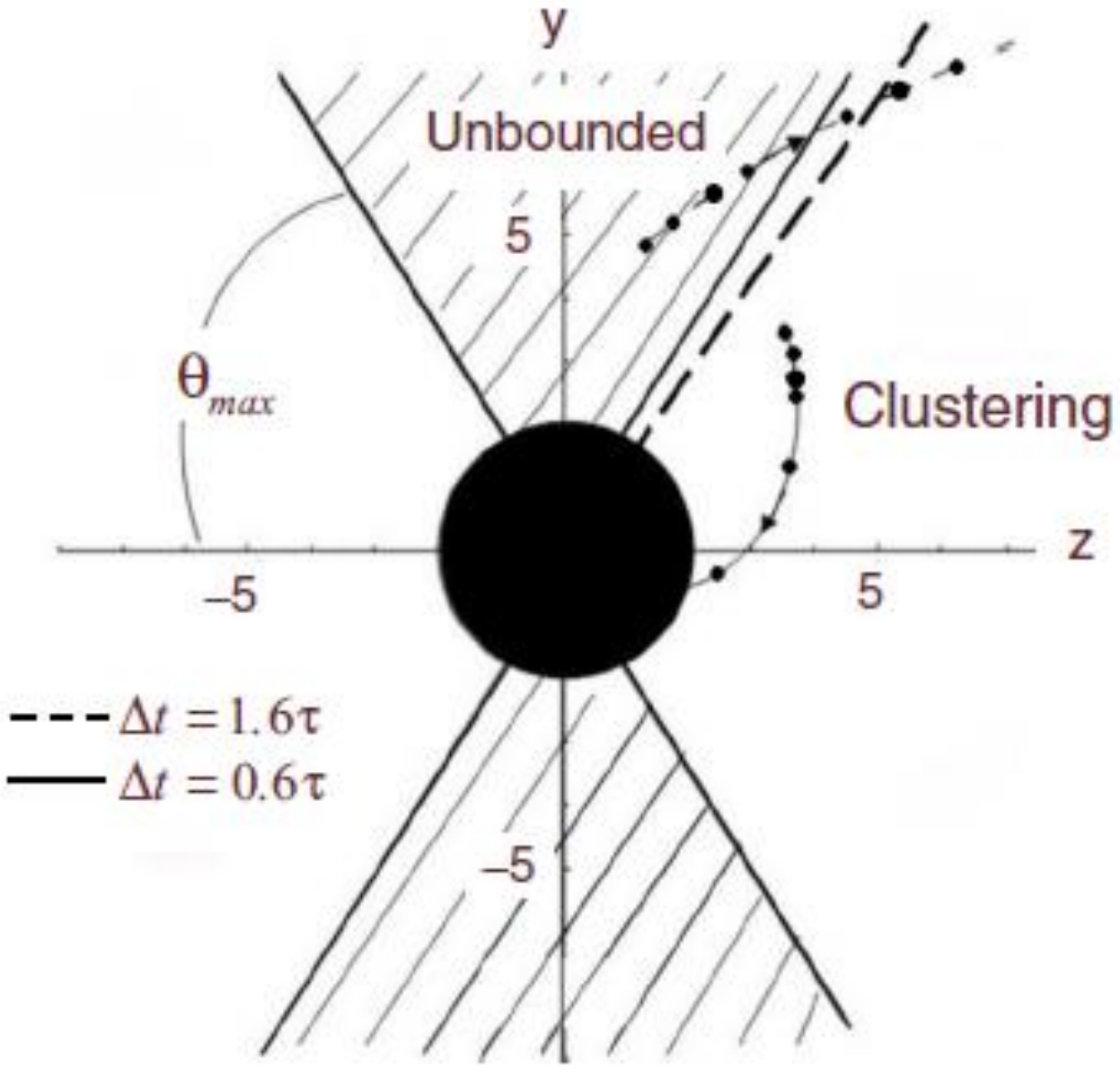

**Figure 2.** (from ref. [7]) Regions of diverging (shaded) and clustering (unshaded) initial position of trajectories in the absence of viscosity. One particle is at the origin while the center of the second particle can be anywhere except for the circle centered at the origin, excluded by the contact repulsive interaction. The axes are in units of the particles radius.

This example shows that light-induced forces may favor cluster formation of neutral particles in suspension. Evidence for such behavior has been experimentally found in several works as cited in ref. [7]

### 3.2 Theoretical analysis for an isotropic particle irradiated by a circularly polarized beam

As known from the corpuscular representation of light, each photon in a left/ right circularly polarized beam of light carries ± $\hbar$ units of angular momentum along the direction of propagation ($\hbar = h/2\pi$ is the Plank's constant). See for example ref. [18].

D. Kiang and K. Young [8] calculate the angular momentum of a photon in a circularly polarized beam by evaluating the energy and angular momentum imparted by a classical plane wave to a point charge considering linear response. Their result is valid for an absorbing particle and no explicit consideration is made to scattering and conservation of linear and angular momentum.

P.L. Marston and J.H. Crichton [19] have shown that an unbounded circularly polarized plane incident electromagnetic wave of frequency $\omega$ propagating in a direction chosen as $z$, transfers angular momentum to an absorbing isotropic sphere of radius $a$, producing a torque with respect to an axis parallel to the direction z that goes through the center of mass (CM) of the particle and is given by:

$$\Gamma_z = I_{inc}\pi a^2 \frac{Q_{abs}}{\omega} \tag{11}$$

where $I_{inc}$ is the intensity of the incident wave and $Q_{abs}$ is the Mie absorption efficiency. According to equation (11) for an idealized lossless dielectric sphere where $Q_{abs}= 0$, no torque would be developed by a circularly polarized electromagnetic wave.

In terms of photons the generation of a torque on a particle through absorption may be interpreted as follows: by angular momentum conservation the extinction of a photon with energy $\hbar\omega$ transfers to the absorbing particle $\hbar$ units of angular momentum in the direction of propagation. The torque generated is just $\hbar$ times the number of photons absorbed by unit of time. For an isotropic spherical particle the torque is only generated by absorption in a form compatible with energy, linear and angular momentum conservation as shown below.

If $F_z$ is the rate of photons incident over the sphere and travelling in the z direction (number of photons per unit of time, each one with energy $\hbar\omega$), $A$ is the photon absorption rate and $G$ is the total rate of scattered photons in all directions, and in assuming elastic scattering, conservation of energy requires,

$$F_z\left(\hbar\omega\right)=A\left(\hbar\omega\right)+G\left(\hbar\omega\right). \tag{12}$$

As known, for an unbounded electromagnetic plane wave, the electric field vector is in a plane perpendicular to the propagation direction, so in our case this field has only components oscillating in the $x$ and $y$ directions which produce oscillating electric dipole vectors also in these directions. By symmetry and taking into account that the radiated intensity of an oscillating dipole is null in a direction parallel to its direction and maximum for a perpendicular direction, one may expect that in the x and y direction there are no scattered photons, and therefore no angular momentum or linear momentum is transferred in these directions.

Considering that each incident photon carries a momentum $\hbar\omega/c$ in the +z direction and denoting $G_z$ as the net rate of photons scattered in the +z direction, conservation of linear momentum requires:

$$F_z\left(\frac{\hbar\omega}{c}\right)=\frac{dP_z}{dt}+G_z\left(\frac{\hbar\omega}{c}\right) \tag{13}$$

where $P_z$ is the linear momentum acquired by the particle in the z direction.

In addition, each incident photon carries an angular momentum $\hbar$ and conservation of angular momentum requires:

$$F_z\left(\hbar\right)=\frac{dL_z}{dt}+G_z\left(\hbar\right) \tag{14}$$

where $L_z$ is the $z$ component of the angular momentum acquired by the particle.

From equations (12) to (14) the torque acting over the particle may be recast as:

$$\Gamma_z = \hbar\left[A + \left(G - G_z\right)\right] \qquad (15)$$

For an isotropic spherical particle of radius *a* comparable with the wavelength λ of the incident radiation, the analysis of the angular distribution of the scattered radiation shows that the scattering is highly peaked in the forward direction [20]. Therefore $G_z = G$ and Eq. (15) gives

$$\Gamma_z = \frac{dL_z}{dt} = A\hbar = \frac{\text{Absorbed Power}}{\omega} \qquad (16)$$

On the other hand, for a small spherical particle such that $a / \lambda << 1$, it can be shown that the scattering may be analyzed in the Rayleigh regime what implies that the effect of scattering may be neglected in comparison with the absorption, so Eq. (15) leads also to the result given by equation (16).

According to equations (11) and (16) the torque on an isotropic and absorbing particle irradiated by an unbounded circularly polarized electromagnetic wave is essentially only due to absorption. Therefore in an idealized lossless dielectric and isotropic sphere no torque would be developed. This analysis may be compared with the predictions of a classical electromagnetic model as follows.

Let us consider a particle centered at the origin, whose polarizability may be considered as isotropic. The particle is subjected to two electric fields applied in perpendicular directions denoted as *x* and *y*, each oscillating with frequency *ω* (see Fig. 3). If the phase difference of the fields is π / 2 the electric field vector may be expressed as:

$$\vec{E}(t) = E_0(\hat{x}\cos\omega t + \hat{y}\sin\omega t) \qquad (17)$$

If the particle complex polarizability is $\alpha = |\alpha| e^{-i\varphi}$ the components of the induced electric dipole are

$$p_x(t) = |\alpha| E_o \cos(\omega t - \varphi),$$
$$p_y(t) = |\alpha| E_o \sin(\omega t - \varphi) \tag{18}$$

The torque with respect to the *z* axis is obtained from

$$\Gamma_z = |\vec{p} \times \vec{E}| = p_x E_y - p_y E_x = |\alpha| E_o^2 \sin\omega t \cdot \cos(\omega t - \varphi) - |\alpha| E_o^2 \cos\omega t \cdot \sin(\omega t - \varphi)$$
$$= |\alpha| E_o^2 \sin\varphi = E_o^2 \operatorname{Im}[\alpha] \tag{19}$$

The time averaged value of the absorbed power by the particle due to the external field is given by [20]

$$\overline{P}_{abs} = \omega E_o^2 \operatorname{Im}[\alpha] \tag{20}$$

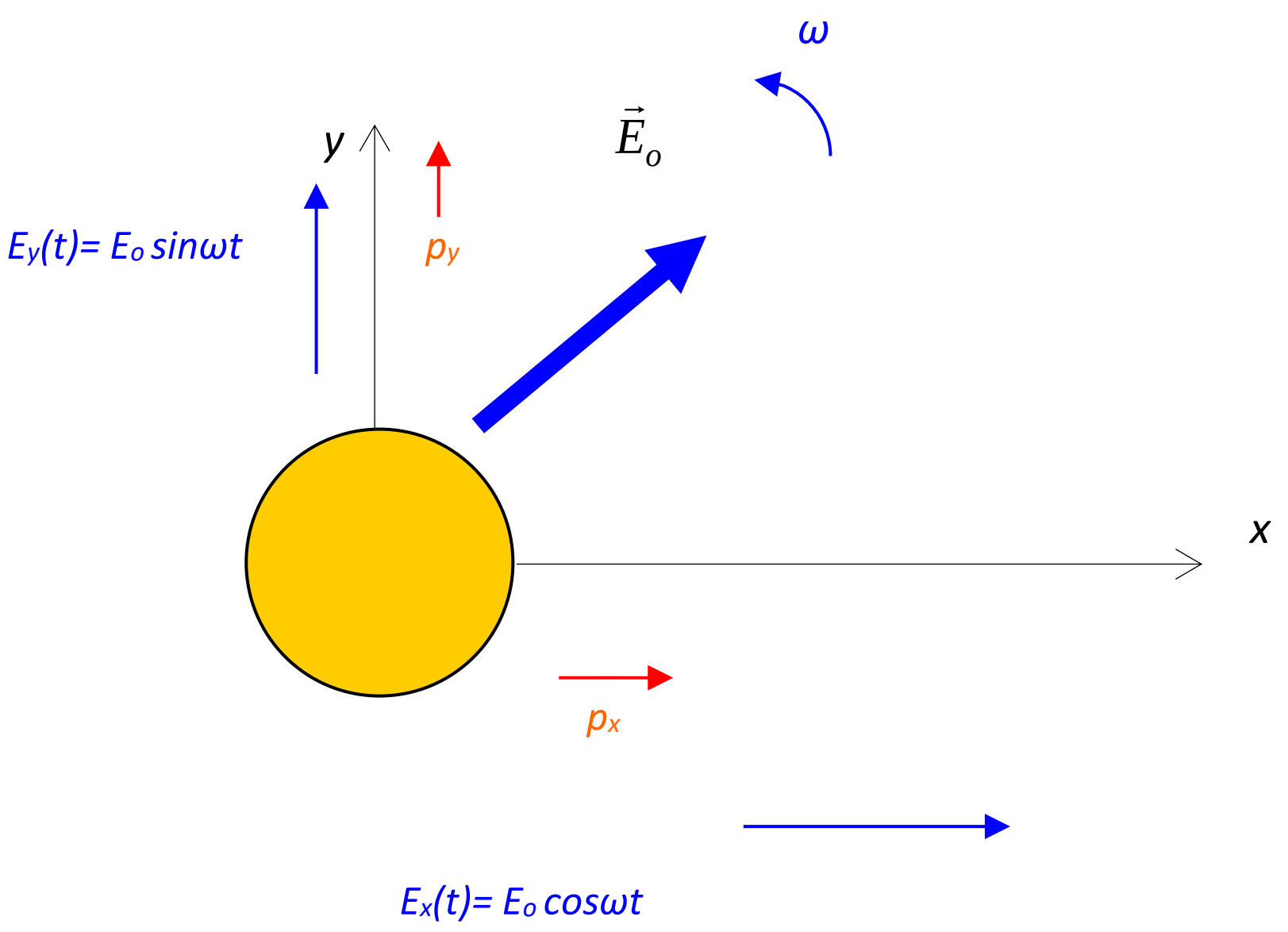

**Figure 3.** Isotropic sphere subjected to an electric field vector rotating counter-clock wise with angular velocity *ω*.

Taking into consideration that the imaginary component of the polarizability is determined only by absorption, this result is in agreement with equation (11). Therefore the predictions obtained by this classical model are consistent with the simplified analysis in terms of photons absorbed by the sphere.

### 3.3 Analysis for an anisotropic and non-symmetrical particle

In general the use of absorbing particles for the transfer of angular momentum by light is impractical due to heating. For that purpose it is better to use weak absorbing particles that are subject to a torque due to scattering [21]. We shall discuss this regime assuming that only elastic scattering takes place when the particle is much smaller than the wavelength of the incident light, and that the scattered radiation may be considered as produced by an oscillating electric dipole [22].

For an anisotropic and absorbing particle there are contributions to the torque due to absorption and scattering. This statement is coherent with experimental results obtained in angular trapping of birefringent particles such as quartz or calcite where polarizabilities are different in two perpendicular directions parallel to their optical axes [23].

The above statement can be compared with the predictions of a fully classical analysis for the interaction of a rotating electric field with an elongated particle such as a linear chain of identical nanoparticles, or a diatomic molecule. Here two symmetry axis may be identified: one axis, denoted as $d$, along a line between the centers of the particles of the cluster and the other, denoted as $q$, in a perpendicular direction, as shown in Fig. 4. The electric field vector rotates in the plane $y$-$x$ (parallel to plane $d$-$q$) with angular velocity $\omega$ and at a given instant its direction is at an angle $\theta_E = \omega t + \theta_{Eo}$ respect to the x axis and at angle $\beta = \omega t - \theta$ respect to the $d$ axis if $\theta_{Eo} = 0$.

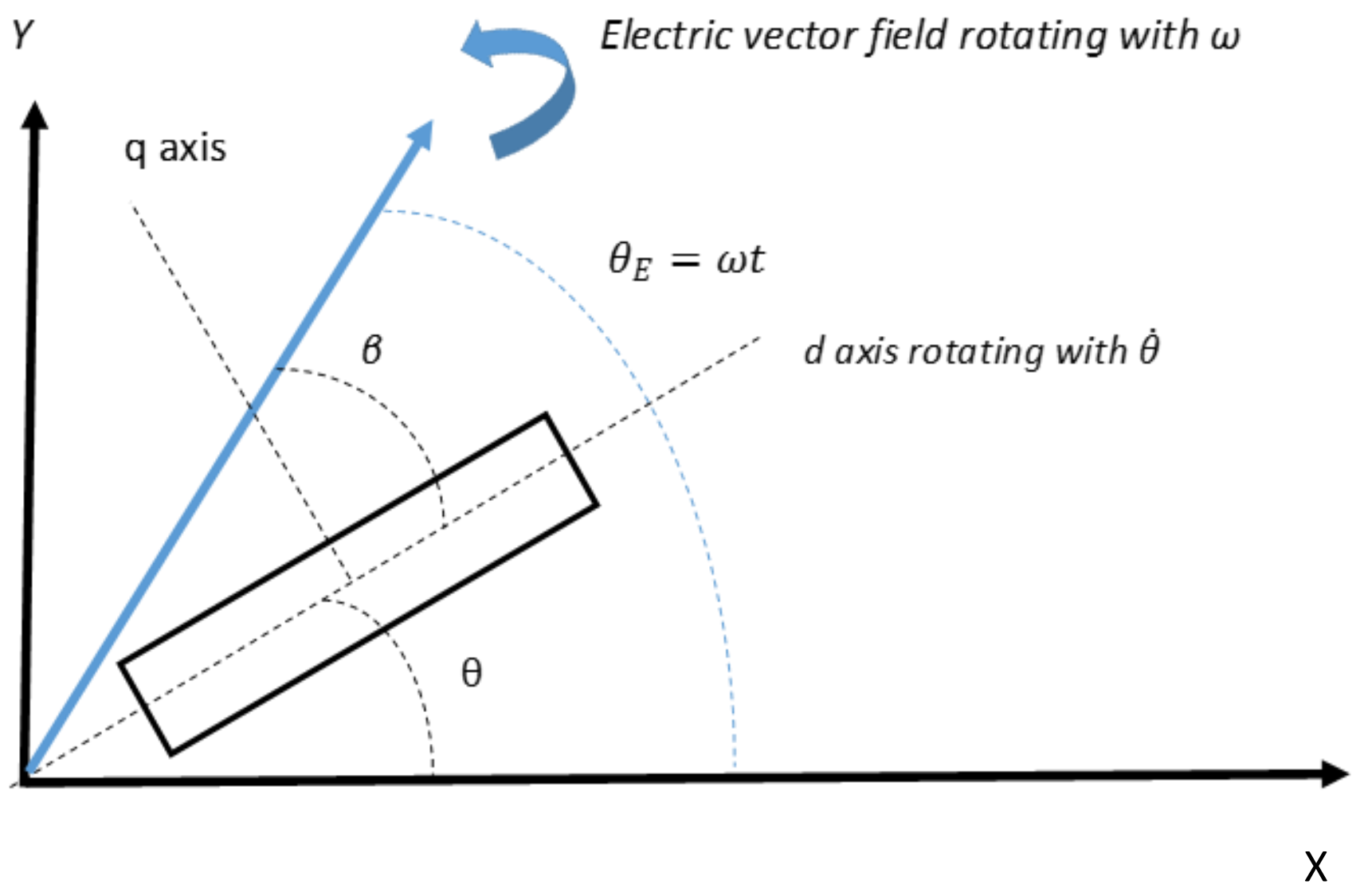

**Figure 4.** Elongated particle subjected to an electric field rotating counter-clock wise with angular velocity ω.

This incident electric field may be expressed as:

$$\vec{E} = E_o(\hat{x} - i\hat{y})e^{i\omega t} \tag{21}$$

The corresponding electric dipoles induced along the *d* and q axis are given by:

$$p_d = |\alpha_d||\vec{E}|\cos\beta\ e^{i(\omega t - \varphi_d)}, \tag{22}$$

$$p_q = |\alpha_q||\vec{E}|\sin\beta\ e^{i(\omega t - \varphi_q)}, \tag{23}$$

x

where $\alpha_d$ and $\alpha_q$ are the complex polarizabilities written as $\alpha_d = |\alpha_d|e^{-i\varphi_d}$ and $\alpha_q = |\alpha_q|e^{-i\varphi_q}$ respectively.

The interaction between these electric dipoles and the electric field vector results in a time averaged value of the torque acting upon the cluster, given by:

$$\Gamma_z = \frac{1}{2}\text{Re}\left[p_d^* E_q\right] - \frac{1}{2}\text{Re}\left[p_q^* E_d\right]$$

$$\Gamma_z = \frac{1}{2}\text{Re}\left[|\alpha_d|\left|\vec{E}\right|\cos\beta e^{-i(\omega t-\varphi_d)}\left|\vec{E}\right|\sin\beta e^{i\omega t}\right] - \frac{1}{2}\text{Re}\left[|\alpha_q|\left|\vec{E}\right|\sin\beta e^{-i(\omega t-\varphi_q)}\left|\vec{E}\right|\cos\beta e^{i\omega t}\right]$$

$$\Gamma_z = \frac{E_o^2}{2}\left\{\text{Re}\left[\alpha_d\right] - \text{Re}\left[\alpha_q\right]\right\}\sin 2\beta \qquad (24)$$

In the last equation we have used the relation $\left|\vec{E}\right|^2 = \vec{E}\cdot\vec{E}^* = 2E_o^2$ .

The instantaneous value of the absorbed power is calculated as:

$$P_{abs}(t) = \text{Re}\left[E_d\right]\text{Re}\left[\frac{dp_d}{dt}\right] + \text{Re}\left[E_q\right]\text{Re}\left[\frac{dp_q}{dt}\right]$$

$$P_{abs}(t) = \text{Re}\left[\left|\vec{E}\right|\cos\beta e^{i\omega t}\right]\text{Re}\left[i\omega|\alpha_d|\left|\vec{E}\right|\cos\beta e^{i(\omega t-\varphi_d)}\right] + \text{Re}\left[\left|\vec{E}\right|\sin\beta e^{i\omega t}\right]\text{Re}\left[i\omega|\alpha_q|\left|\vec{E}\right|\sin\beta e^{i(\omega t-\varphi_q)}\right] \qquad (25)$$

The time averaged value of this power is

$$P_{abs} = \frac{1}{2}\text{Re}\left[\left|\vec{E}\right|\cos\beta e^{-i\omega t} i\omega|\alpha_d|\left|\vec{E}\right|\cos\beta e^{i(\omega t-\varphi_d)}\right] + \frac{1}{2}\text{Re}\left[\left|\vec{E}\right|\sin\beta e^{-i\omega t} i\omega|\alpha_q|\left|\vec{E}\right|\sin\beta e^{i(\omega t-\varphi_q)}\right] \qquad (26)$$

The above expression leads to the result:

$$P_{abs} = \frac{\omega E_o^2}{2}\left\{\text{Im}\left[\alpha_d\right] + \text{Im}\left[\alpha_q\right] + \left(\text{Im}\left[\alpha_d\right] - \text{Im}\left[\alpha_q\right]\right)\cos 2\beta\right)\right\} \qquad (27)$$

Therefore the resulting torque with respect to the CM of the cluster is obtained by adding the contributions of the alignment torque, Eq. (24), and the torque due to absorption, Eq.(27), leading to

$$\bar{\Gamma} = \frac{P_{abs}}{\omega} + \frac{E_o^2}{2}\left(\text{Re}\left[\alpha_d\right] - \text{Re}\left[\alpha_q\right]\right)\sin 2\beta$$

$$\bar{\Gamma} = \frac{E_o^2}{2}\left\{\text{Im}\left[\alpha_d\right] + \text{Im}\left[\alpha_q\right] + \left(\text{Im}\left[\alpha_d\right] - \text{Im}\left[\alpha_q\right]\right)\cos 2\beta\right\} + \frac{E_o^2}{2}\left\{\text{Re}\left[\alpha_d\right] - \text{Re}\left[\alpha_q\right]\right\}\sin 2\beta \qquad (28)$$

According to equation (28) a lossless anisotropic particle still acquires a torque due to scattering of circularly polarized light. The interpretation of eq. (28) is as follows:

1. The absorbed power corresponds to photon absorption, each one carrying angular momentum transferred to the particle and contributing to produce a torque according to eq. (11). This torque corresponds to the first term of Eq. (28) and is related to the imaginary part of the polarizabilities.

2. The second term represents an alignment torque component due to the anisotropy in the polarization: one of the axes of the particle is more easily polarized than the other. The same effect occurs in a diatomic molecule where there is an interaction potential between the electric dipole moments induced in directions parallel and perpendicular to the axis of the molecule, which is of the form $-U_0 \cos^2\beta$ where $U_0$ is proportional to the difference between the polarizabilities in these directions [24].

3. Objects or particles of elongated shape are birefringent and therefore change the polarization state of a light beam, so that it would be expected that a torque independent of the orientation of the particle with respect to the beam axis will be produced, providing the possibility of rotation of the particle at a constant rate [25].

According to Eq. (28), a lossless anisotropic particle interacting with circularly polarized light acquires a torque due to angular momentum transfer. If the particle is in a liquid with angular drag coefficient $\gamma$, the dynamical equation is

$$\frac{E_O^2}{2}\left\{\alpha_d - \alpha_q\right\}\sin 2\beta = J\frac{d^2\theta}{dt^2} + \gamma\frac{d\theta}{dt}, \tag{29}$$

where $J$ is the moment of inertia of the particle with respect to an axis perpendicular to the plane $x$-$y$ and passing through its center of gravity. As shown in [16] in steady state conditions the particle reaches an angular velocity equal to that of the applied electric rotating field provided that the following condition is satisfied:

$$\sin 2\beta = \frac{2\gamma\omega}{E_o^2\left(\alpha_d - \alpha_q\right)} \leq 1 \tag{30}$$

This condition corresponds to a synchronous behavior and this device is called a synchronous nanomotor (SN), considering that the dimensions of the considered particle are in the scale of nanometers.

If the condition corresponding to Eq. (30) is not fulfilled the particle cannot rotate synchronously with the applied field and angle $\beta$ is a periodic function of time. If $d^2\theta/dt^2$ may be neglected (acceptable approach for a nanoparticle) this angle may be obtained by integration of Eq. (29) as

$$\gamma\omega\ \tan\beta(t) = A + \sqrt{\gamma^2\omega^2 - A^2}\ \tan\left[\sqrt{\omega^2 - (A/\gamma)^2}\ t - \tan^{-1}\left[\frac{A}{\sqrt{\gamma^2\omega^2 - A^2}}\right]\right] \tag{31}$$

where $A = E_o^2\left(\alpha_d - \alpha_q\right)/2$. This equation has been obtained arbitrarily assuming that $\beta(0) = 0$, meaning that the excitation has been turned on when the corresponding electric field vector is parallel to the d axis shown in Fig. 4.

We can gain insight into the solution of the nonlinear differential equation (29) for fields below the threshold by taking the limit of a very weak field. To first order, the left-hand side may then be neglected. The solution is an exponential decay where $\theta_1(t) = Ae^{-t/\tau} + B$ with $A$ and $B$ constants and $\tau = \gamma/J$. In the next order of approximation we replace $\theta_1(t)$ on the lefthand side of Eq. (29) and realize that for times greater than $\tau$ this equation has a solution that oscillates with angular frequency twice the external angular frequency.

The minimum value of the intensity of the applied electromagnetic field required for a constant rotation frequency $f_r$ of the particle (in condition of synchronism this frequency is equal to that of the rotating electric field), is obtained from Eq. (30) considering that when steady state conditions are reached, $\ddot{\theta} = 0$ and $\sin 2\beta \leq 1$, obtaining

$$I_{L\min} = \frac{2\pi\ \gamma f_r\ c\sqrt{K_m}}{\alpha_d - \alpha_q}, \tag{32}$$

where $K_m$ is the high frequency limit dielectric constant of the medium in which the particle is immersed.

### 3.4. Forces and torque between interacting particles including higher order multipoles

In the above sections we have considered only the dipolar approach for determining the dipole moments induced in each particle and the corresponding forces and torques. As known [26], for a dilute system of spherical particles with separations of the order of three or more times their radii, the accuracy of the results given by the dipole approximation is acceptable and the effect of higher order multipoles may be neglected. In such case and if only two particles are present, the electric force between them may be calculated from equations (6) and (7) of section 3.1. For separations bettween particles less than three particle radii, the effect of higher multipoles must be included. Reference [27] presents the Normal Mode Theory which allows to include this effect for arrays of spherical particles.

In this subsection we describe a general model for determining the force and torque over each particle of a set of spherical nanoparticles placed at random positions, giving expressions for the time-averaged interaction forces and torques for any separation between the particles. The approach is based in refs. [26-28] and all the expressions are in Gaussian units.

We consider a system of $N$ spherical nanoparticles of radii $a_i$ placed at random positions in a non- absorptive dielectric medium, and excited by a laser field of angular frequency $\omega$, as shown in Fig. 5. The particles are uncharged and their response to a local electric field is characterized by a complex dielectric function $\varepsilon(\omega)$.

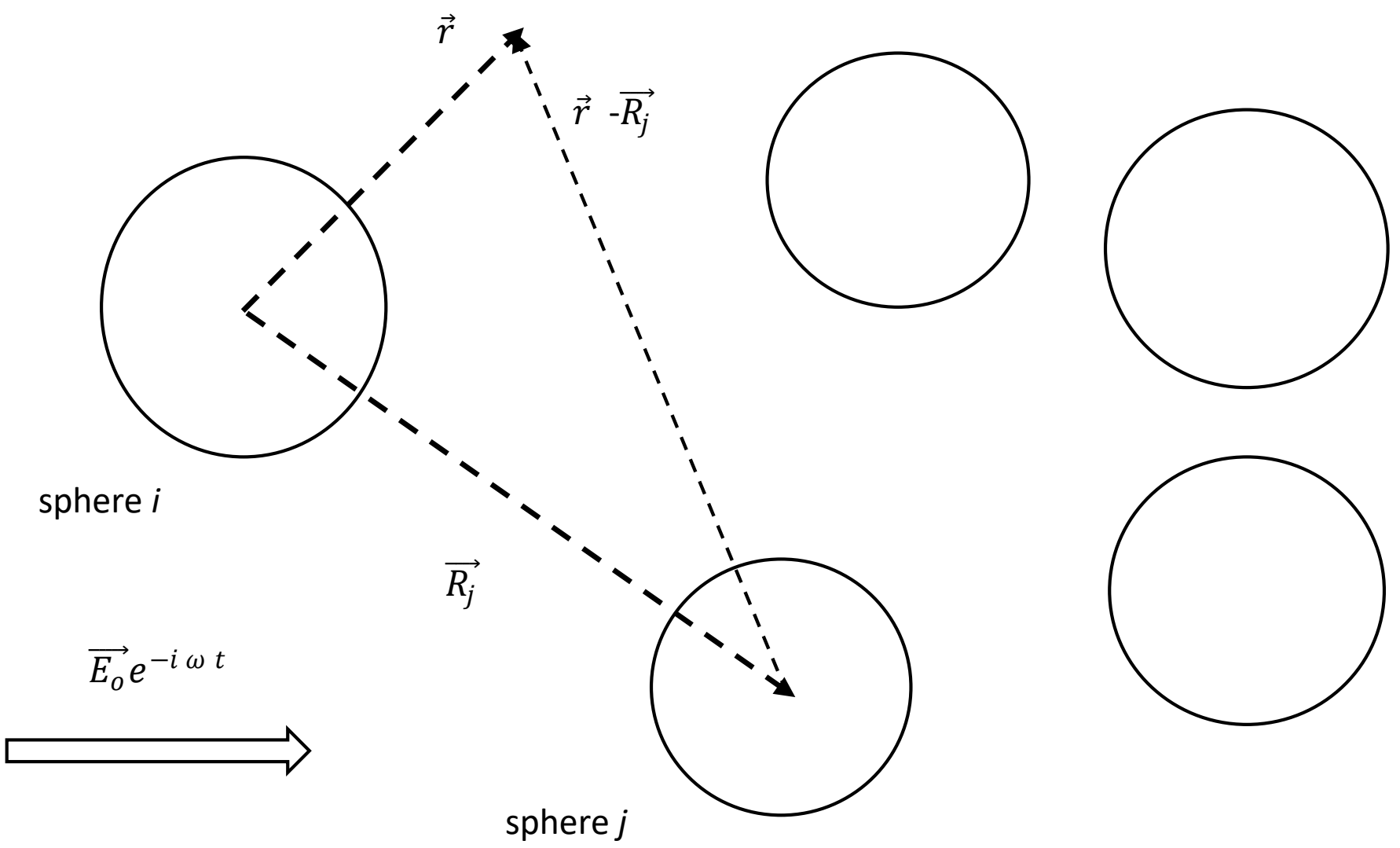

**Figure 5.** Array of spherical particles subjected to an oscillating electric field.

It is assumed that the size of the system is much smaller than the wavelength of the incident light, thus allowing to use a non-retarded theory in which the local electric field may be expressed as the gradient of an electric potential which does not exert any force on the uncharged spheres. The external field induces a dipole on each particle, which in turn excites multipole moments on every other sphere owing to the nonuniformity of the electric field it produces at each particle site.

For the particle *i*, if the internal induced charge density is $\rho_i(\vec{r})$ the multipole of order *l*, *m* is given by [22]

$$q_{lmi} = \int \rho_i(\vec{r}) r^l \, Y^*_{lm}(\theta, \varphi) d^3r \tag{33}$$

where $Y_{lm}(\theta, \varphi)$ is the corresponding spherical harmonic and the integral covers only the volume of the particle *i*.

Choosing a reference frame with the origin at the center of particle $i$ as shown in Fig. 5, the electric potential at a point $\vec{r}=(r,\theta,\varphi)$ in the medium is given by the following expression

$$V(\vec{r})=\sum_{l=1}^{\infty}\sum_{m=-l}^{+l}\frac{4\pi}{2l+1}q_{lmi}\frac{Y_{lm}(\theta,\varphi)}{r^{l+1}}+\sum_{l=1}^{\infty}\sum_{m=-l}^{+l}\sum_{j\neq i}\frac{4\pi}{2l+1}q_{lmj}\frac{Y_{lm}(\bar{\theta}_j,\bar{\varphi}_j)}{\bar{R}_j^{\,l+1}}. \tag{34}$$

Sphere $j$ is at $\vec{R}_j$ and $\vec{r}-\vec{R}_j\equiv(\bar{R}_j,\bar{\theta}_j,\bar{\varphi}_j)$ is the position vector of the observation point with respect to spherical particle $j$. The first term of Eq. (33) is the contribution to the potential at the point placed at $\vec{r}$ coming from the multipole $q_{lmi}$, while the second term is the contribution from the multipoles $q_{lmj}$ induced in the other particles $j\neq i$. There is an electromagnetic coupling between the multipoles expressed as

$$q_{lmi}=-\frac{la_i^{2l+1}}{4\pi\left(\frac{1}{\epsilon-1}+\frac{l}{2l+1}\right)}\left[V_{lm}^{ext}(i)+\sum_{l',m',j}(-1)^{l'}A_{l,m,i}^{l',m',j}\,q_{l',m',j}\right], \tag{35}$$

where $A_{lmi}^{l'm'j}$ is the coupling coefficient between $q_{lmi}$ and $q_{l'm'j}$ $(i\neq j)$ given by

$$A_{lmi}^{l'm'j}=(-1)^{m'}\frac{\left[Y_{l+l',m-m'}(\theta_i-\theta_j,\varphi_i-\varphi_j)\right]^*}{\left|\vec{R}_i-\vec{R}_j\right|^{l+l'+1}}\left[\frac{(4\pi)^3(l+l'+m-m')!(l+l'-m+m')!}{(2l+1)(2l'+1)(2l+2l'+1)\;(l+m)!(l-m)!(l'+m')!(l'-m')!}\right]^{1/2}$$

(36)

$V_{lm}^{ext}(i)$ is a coefficient of expansion of the potential associated to the applied electric field. If the external potential corresponds to a uniform electric field in the z-direction, $\vec{E}^{ext}(\vec{r})=E_z^{ext}\,\hat{z}$, it can be shown that $V_{lm}^{ext}=-\sqrt{4\pi/3}E_z^{ext}\delta_{l1}\delta_{m0}$. Furthermore $\epsilon$ is the dielectric function of the particles and $a_i$ is the radius of spherical particle $i$.

Equation (35) constitutes an infinite set of equations, but it must be considered that the couplings become weaker as the indices $l,l'$ become larger, so that it is a good approach to cut off the infinite sequence $l=0,1,2,\ldots$ at some $l_{max}=L$.

The mathematical procedure in order to get average forces between particles in this model is summarized as follows:

- Rewrite the potential of Eq. (34) in terms of coordinates $(r, \theta, \phi)$ respect to the center of particle *i*.

- Obtain the local field acting over particle *i* as the gradient of the new expression for the potential.

-Using Coulomb's law expressed in terms of multipoles, obtain the average force over particle *i*.

<u>Next we follow the procedure step by step.</u>

Vectors $\vec{r}$ and $\vec{R}_j$ in Eq. (34) are uncoupled by using identities. Then we add the potential $V^{ext}$ due to the external field, and equation (34) becomes

$$V(\vec{r}) = \sum_{l,m} \frac{4\pi}{2l+1} q_{lmi} \frac{Y_{lm}(\theta,\varphi)}{r^{l+1}} + \sum_{l,m} b_{lmi} Y_{lm}(\theta,\varphi) r^l + V^{ext}. \tag{37}$$

The average force over particle *i* is obtained from

$$\langle \vec{F}_i \rangle = \frac{1}{2} Re \int \rho_i^*(\vec{r}) \vec{E}(\vec{r}) d^3\vec{r}, \tag{38}$$

where $\rho_i$ is the induced charge density of particle *i* and $\vec{E}(\vec{r})$ is the local electric field at particle i, due to the external sources and the other particles in the ensemble. It is obtained from $\vec{E}(\vec{r}) = -\nabla V_i(\vec{r})$ and

$$V_i(\vec{r}) = \sum_{lm} b_{lmi}\, r^l Y_{lm}(\theta,\varphi) + V^{ext}(\vec{r}). \tag{39}$$

Inside the *i*-th sphere the potential is given by

$$V_{\text{int},i}(r) = \sum_{l,m} c_{lmi} \left(\frac{r}{a_i}\right)^l Y_{lm}(\theta,\varphi) \tag{40}$$

In this last equation, $c_{lmi}$ is a coefficient determined from the condition of continuity of the potential at the surface of particle *i*, $V_{int,i}(a_i) = V_i(a_i)$.

It must be considered that in Eq. (38) $\rho_i(\vec{r})$ is written as a volume charge density, and the corresponding surface charge density induced on sphere *i* is determined by

$$\sigma_i = \frac{1}{4\pi}\left\{ \partial V_{\mathrm{int},i} / \partial r \,\Big|_{r=a_i} - \partial V_i / \partial r \,\Big|_{r=a_i} \right\}. \tag{41}$$

In order to obtain explicit expressions for the components of the force, in Eq. (40) the spherical harmonics are written in terms of Legendre functions by using the relation $Y_{lm}(\theta,\varphi) = \sqrt{(2l+1)(l-m)\,! \,/\, \left[4\pi(l+m)\,!\right]}\,P_l^m(\cos\theta)e^{im\varphi}$.

Then Eq. (39) may be recast as

$$V_i(\vec{r}) = \sum_{lm} D_{lmi} r^l P_l^m(\cos\theta)e^{im\varphi} \tag{42}$$

where

$$D_{lmi} = V_{lmi}\sqrt{(2l+1)(l-m)\,! \,/\, \left[4\pi(l+m)\,!\right]}. \tag{43}$$

Therefore the spherical components of the local electric field are

$$E_r = -\frac{\partial V_i(\vec{r})}{\partial r} = -\sum_{lm} D_{lmi} l\, r^{l-1} P_l^m(\cos\theta)e^{im\varphi} \tag{44}$$

$$E_\theta = -\frac{1}{r}\frac{\partial V_i(\vec{r})}{\partial\theta} = -\sum_{lm} D_{lmi} r^{l-1}\frac{\partial}{\partial\theta}P_l^m(\cos\theta)e^{im\varphi} = \sum_{lm} D_{lmi}\, r^{l-1}\sqrt{1-\xi^2}\,\frac{\partial}{\partial\xi}P_l^m(\xi)e^{im\varphi} \tag{45}$$

$$E_\phi = -\frac{1}{r\sin\theta}\frac{\partial V_i(\vec{r})}{\partial\phi} = -\sum_{lm} D_{lmi} r^{l-1}\frac{1}{\sqrt{1-\xi^2}}P_l^m(\xi)\,im e^{im\varphi} \tag{46}$$

In Eqs. (45) and (46) we have defined $\xi = \cos\theta$. The corresponding Cartesian components of the electric field are given by

$$E_x = E_r \sin\theta\cos\phi + E_\theta \cos\theta\sin\phi - E_\phi \sin\phi \tag{47}$$

$$E_y = E_r \sin\theta\sin\phi + E_\theta \cos\theta\cos\phi + E_\phi \cos\phi \tag{48}$$

$$E_z = E_r \cos\theta - E_\theta \sin\theta \tag{49}$$

It is useful to calculate linear combinations of $E_x$ and $E_y$ defined as

$$E_+ = E_x + iE_y \tag{50}$$

$$E_- = E_x - iE_y \tag{51}$$

From the above equations one obtains

$$E_+ = -\sum_{lm} D_{lmi} r^{l-1} \left[ l\sqrt{1-\xi^2} P_l^m(\xi) - \xi\sqrt{1-\xi^2}\frac{\partial}{\partial\xi} P_l^m(\xi) - \frac{m}{\sqrt{1-\xi^2}} P_l^m(\xi) \right] e^{i(m+1)\varphi} \tag{52}$$

The relations

$$\left(1-\xi^2\right)\partial P_l^m(\xi)/\partial\xi = (l+m)P_{l-1}^m(\xi) - l\xi P_l^m(\xi) \tag{53}$$

$$(l-m)P_l^m(\xi) - \xi(l+m)P_{l-1}^m(\xi) = \sqrt{1-\xi^2}\, P_{l-1}^{m+1}(\xi) \tag{54}$$

lead to

$$E_+ = -\sum_{lm} D_{lmi} r^{l-1}\, P_{l-1}^{m+1}(\xi) e^{i(m+1)\varphi} \tag{55}$$

From Eq. (A 27) Appendix A of ref. [28] it can be obtained

$$E_+ = -\sum_{lm} V_{lmi} r^{l-1} \sqrt{\frac{2l+1}{2l-1}(l-m)(l-m-1)}\, Y_{l-1,m+1}(\theta,\varphi) \tag{56}$$

Similarly,

$$E_- = -\sum_{lm} D_{lmi} r^{l-1} \left[ l\sqrt{1-\xi^2} P_l^m(\xi) - \xi\sqrt{1-\xi^2}\frac{\partial}{\partial\xi} P_l^m(\xi) + \frac{m}{\sqrt{1-\xi^2}} P_l^m(\xi) \right] e^{i(m+1)\varphi}\,. \tag{57}$$

The recurrence relation $\xi P_{l-1}^m(\xi) - P_l^m(\xi) = (l+m-1)P_{l-1}^{m-1}(\xi)$ can be used to find that

$$E_- = \sum_{lm} V_{lmi} r^{l-1} \sqrt{\frac{2l+1}{2l-1}(l-m)(l+m-1)}\, Y_{l-1,m-1}(\theta,\varphi) \tag{58}$$

To obtain $E_z$ we use the relation $\frac{(1-\xi^2)\partial P_l^m(\xi)}{\partial\xi} = (l+m)P_{l-1}^m(\xi) - l\xi P_l^m(\xi)$ to give

$$E_z = -\sum_{lm} V_{lmi} r^{l-1} \sqrt{\frac{2l+1}{2l-1}(l-m)(l+m)}\; Y_{l-1,m}(\theta,\varphi)\,. \tag{59}$$

The Cartesian components of the time-averaged force acting upon sphere *i* are then given by

$$\langle F_{ix}\rangle = \frac{1}{2}\mathrm{Re}\int \rho_i^*(\vec{r})E_x d^3\vec{r} \;=\; \frac{1}{2}\mathrm{Re}\int \rho_i^*(\vec{r})\frac{1}{2}(E_+ + E_-)d^3\vec{r}$$

$$= -\frac{1}{4}\mathrm{Re}\int \rho_i^*(\vec{r})\sum_{lm} V_{lmi} r^{l-1}\sqrt{\frac{2l+1}{2l-1}}\left[\sqrt{(l-m)(l-m-1)}Y_{l-1,m+1}(\theta,\varphi) - \sqrt{(l+m)(l+m-1)}Y_{l-1,m-1}(\theta,\varphi)\right]d^3\vec{r}$$

$$= -\frac{1}{4}\mathrm{Re}\sum_{lm} V_{lmi}\sqrt{\frac{2l+1}{2l-1}}\left[\sqrt{(l-m)(l-m-1)}q^*_{l-1,m+1,i} - \sqrt{(l+m)(l+m-1)}q^*_{l-1,m-1,i}\right]$$

(60)

$$\left\langle F_{iy}\right\rangle = \frac{1}{2}\mathrm{Re}\int \rho_i^*(\vec{r})E_y\, d^3\vec{r} = \;\frac{1}{2}\mathrm{Re}\int \rho_i^*(\vec{r})\frac{i}{2}\left(-E_+ + E_-\right)d^3\vec{r}$$

$$= \frac{1}{2}\mathrm{Re}\int \rho_i^*(\vec{r})\frac{i}{2}\sum_{lm} V_{lmi}\; r^{l-1}\sqrt{\frac{2l+1}{2l-1}}\left[\sqrt{(l-m)(l-m-1)}Y_{l-1,m+1}(\theta,\varphi) + \sqrt{(l+m)(l+m-1)}Y_{l-1,m-1}(\theta,\varphi)\right]\, d^3r$$

$$= \frac{1}{4}\mathrm{Re}\left\{i\sum_{lm} V_{lmi}\;\sqrt{\frac{2l+1}{2l-1}}\left[\sqrt{(l-m)(l-m-1)}\; q^*_{l\text{-}1,m+1,i} + \sqrt{(l+m)(l+m-1)}\; q^*_{l\text{-}1,m\text{-}1,i}\right]\right\} \quad (61)$$

$$\left\langle F_{iz}\right\rangle = \frac{1}{2}\mathrm{Re}\int \rho_i^*(\vec{r})E_z d^3\vec{r} = -\frac{1}{2}\mathrm{Re}\int \rho_i^*(\vec{r})\sum_{lm} V_{lmi} r^{l-1}\sqrt{\frac{2l+1}{2l-1}(l-m)(l+m)}Y_{l-1,m}(\theta,\varphi)d^3\vec{r}$$

$$= -\frac{1}{2}\mathrm{Re}\sum_{lm} V_{lmi}\;\sqrt{\frac{2l+1}{2l-1}(l-m)(l+m)}\; q^*_{l\text{-}1,m,i} \quad (62)$$

The coefficients $V_{lmi}$ and $q_{lmi}$ are related by $q_{lmi} = -(2l+1)\alpha_{li}V_{lmi}/4\pi$ where $\alpha_{li}$ is the multipole polarizability of the sphere *i* given by

$$\alpha_{li} = \frac{l(\varepsilon-1)}{l(\varepsilon+1)+1}\; a_i^{2l+1} \quad (63)$$

Using these relations we then obtain the time-averaged force over sphere *i* as

$$\langle F_{ix}\rangle = \pi\,\mathrm{Re}\sum_{lm}(\alpha_{li})^{-1}\left[\sqrt{\frac{(l-m)(l-m-1)}{(2l+1)(2l-1)}}q_{lmi}q^*_{l-1,m+1,i} - \sqrt{\frac{(l+m)(l+m-1)}{(2l+1)(2l-1)}}q_{lmi}q^*_{l-1,m-1,i}\right] \quad (64)$$

$$\langle F_{iy}\rangle = -\pi\,\mathrm{Re}\,i\sum_{lm}(\alpha_{li})^{-1}\left[\sqrt{\frac{(l-m)(l-m-1)}{(2l+1)(2l-1)}}q_{lmi}q^*_{l-1,m+1,i} + \sqrt{\frac{(l+m)(l+m-1)}{(2l+1)(2l-1)}}q_{lmi}q^*_{l-1,m-1,i}\right] \quad (65)$$

$$\langle F_{iz}\rangle = 2\pi\,\mathrm{Re}\sum_{lm}(\alpha_{li})^{-1}\sqrt{\frac{(l-m)(l+m)}{(2l+1)(2l-1)}}\; q_{lmi}q^*_{l-1,m,i} \quad (66)$$

It is worth noting that the components of the force over each particle are expressed in terms just of its own induced multipoles, electromagnetically coupled with the polarization charge of the rest of the ensemble. Varying the corresponding positions of the particles modifies this coupling and therefore, the forces. In appendix B is explained how this model can be applied for obtainig the electrical force between two polarized spherical particles analyzing the limit as the separation increases.

Turning now the attention to the torques as shown in ref. [28] the time-averaged torque on the sphere $i$ due to the local field $\vec{E}$ is given by

$$\langle \vec{\tau}_i \rangle = \frac{1}{2} \mathrm{Re} \int \rho_i^*(\vec{r}) \vec{r} \times \vec{E} \, d^3\vec{r} \tag{67}$$

Work similar to that done above for the forces (see Appendix B of ref. [28]) leads to the time-averaged torque components

$$\langle \tau_{ix} \rangle = \mathrm{Re} \sum_l \frac{2\pi i}{(2l+1)\alpha_{li}} \mathrm{Re}\, S_{li} \,, \tag{68}$$

$$\langle \tau_{iy} \rangle = \mathrm{Re} \sum_l \frac{2\pi i}{(2l+1)\alpha_{li}} \mathrm{Im}\, S_{li} \,, \tag{69}$$

$$\langle \tau_{iz} \rangle = \mathrm{Re} \sum_{lm} \frac{2\pi i m}{(2l+1)\alpha_{li}} q_{lmi} q_{lmi}^* \,, \tag{70}$$

where $S_{li} = \sum_{m=-l}^{l-1} \sqrt{(l-m)(l+m+1)} q_{lmi} q_{l,m+1,i}^*$ and $\alpha_{li}$ is the multipole polarizability of the sphere $i$ given by

$$\alpha_{li} = \frac{l(\varepsilon - 1)}{l(\varepsilon + 1) + 1} a_i^{2l+1} \tag{71}$$

It is clear from Eqs. (68) – (70) that if the system has no dissipation, i.e. if $\alpha_{li}$ is real, the torque is zero. Therefore we conclude that in general torque arises in such systems from dissipative electromagnetic interactions.

Even if there is dissipation, the torque may be suppressed by special symmetries. Such is the case for a linear array subject to a uniform electric field parallel to the line joining their centers. By choosing the z-axis to be aligned with this line only modes with *m=0* are excited leading to zero torque, as may be easily verified from the structure of the above equations. A similar situation occurs if the applied electric field lies in the *xy* plane since in this case only modes with $m=\pm 1$ are excited and the torque is again zero since in this case the sum through *m* in Eq. (70) with values $\pm 1$ vanishes considering that $|q_{l,1,i}|^2 = |q_{l,-1,i}|^2$ (from ref. [22] it can be shown that $q_{l,-1,i} = -q^*_{l,1,i}$).

A torque does arise in such arrays also if they are subject to a rotating electric field on the *xy* plane. The field may be written as $\vec{E} = E_0\left(\pm\hat{x} - i\hat{y}\right)e^{i\omega t}$ and the corresponding coefficients of expansion of the potential are either $V_{1,+1} = \sqrt{2\pi/3}E_0\left(1-i\right)$ or $V_{1,-1} = \sqrt{2\pi/3}E_0\left(-1-i\right)$ depending of the sense of rotation of the electric field vector given by the sign of the *x* component [22]. Correspondingly the excited modes are either $m=+1$ or $m=-1$ (only one of them) and from eq. (70) it follows that a torque may appear. In fact, from Eq. (70) it can be shown that for this case the time-average of the *z* component of the torque is given by

$$\left\langle \tau_{i\,z} \right\rangle = \frac{2\pi m}{a^{2l+1}} \sum_l \frac{l\,\mathrm{Im}(\varepsilon)}{\left[l\left(\mathrm{Re}(\varepsilon)-1\right)\right]^2 + [l\,\mathrm{Im}(\varepsilon)]^2} \left|q_{lmi}\right|^2, \tag{72}$$

For the special case of a single sphere in a rotating external field the torque obtained from this expression is in agreement with references [16] and [19]. The physical origin of such a torque is conservation of angular momentum. The rotating field carries angular momentum, which is transferred to the particles when absorption takes place causing them to experience a spinning torque.

In addition, it must be considered that as noted in Ref. [29] when a linearly polarized field is not aligned with a symmetry axis of a linear array such as a pair, the local field at each particle site has a rotating component, and the same argument applies. In fact, reference [29] shows that when two spherical particles arranged such that a line joining their centres makes an angle $\theta$ with respect to the direction of an applied electric field, the induced electric dipoles establishes a local rotating electric field producing a rotation of these particles.

As an additional comment before closing this section it is worth to mention that the normal modes theory can be applied to the analysis of several other situations besides the calculation of forces and torques here presented.

For instance, as shown in Ref. [30] it can be applied to the analysis of the the effect of multipolar couplings among brine-filled pores in the low-frequency dielectric anomaly of porous media, which gives rise to a large dielectric response having a great impact over optical properties such as electromagnetic absorption.

Other application is for the analysis of the surface enhanced Raman scattering (SERS) which is a sensitive and selective spectroscopic technique for molecular detection and identification. SERS requires the excitation of a localized surface plasmon resonance (LSPR) by an incident electromagnetic wave and the magnitude of the Raman scattering intensity is significantly enhanced when the molecules are placed in the local LSPR fields [31-32]. This has medical applications such as the analysis of samples with a complex biochemical composition, detection of tumours and imaging.

## 4. Synchronous rotation of a ferromagnetic nanowire

We consider an active particle with a permanent magnetic dipole under the action of a rotating magnetic field. Several regimes of the particle motion in a rotating magnetic field are established: synchronous motion at frequencies below the critical and back and forth motion at frequencies above it. The magnetically driven rotation of a single nanowire is shown schematically in Fig. 6.

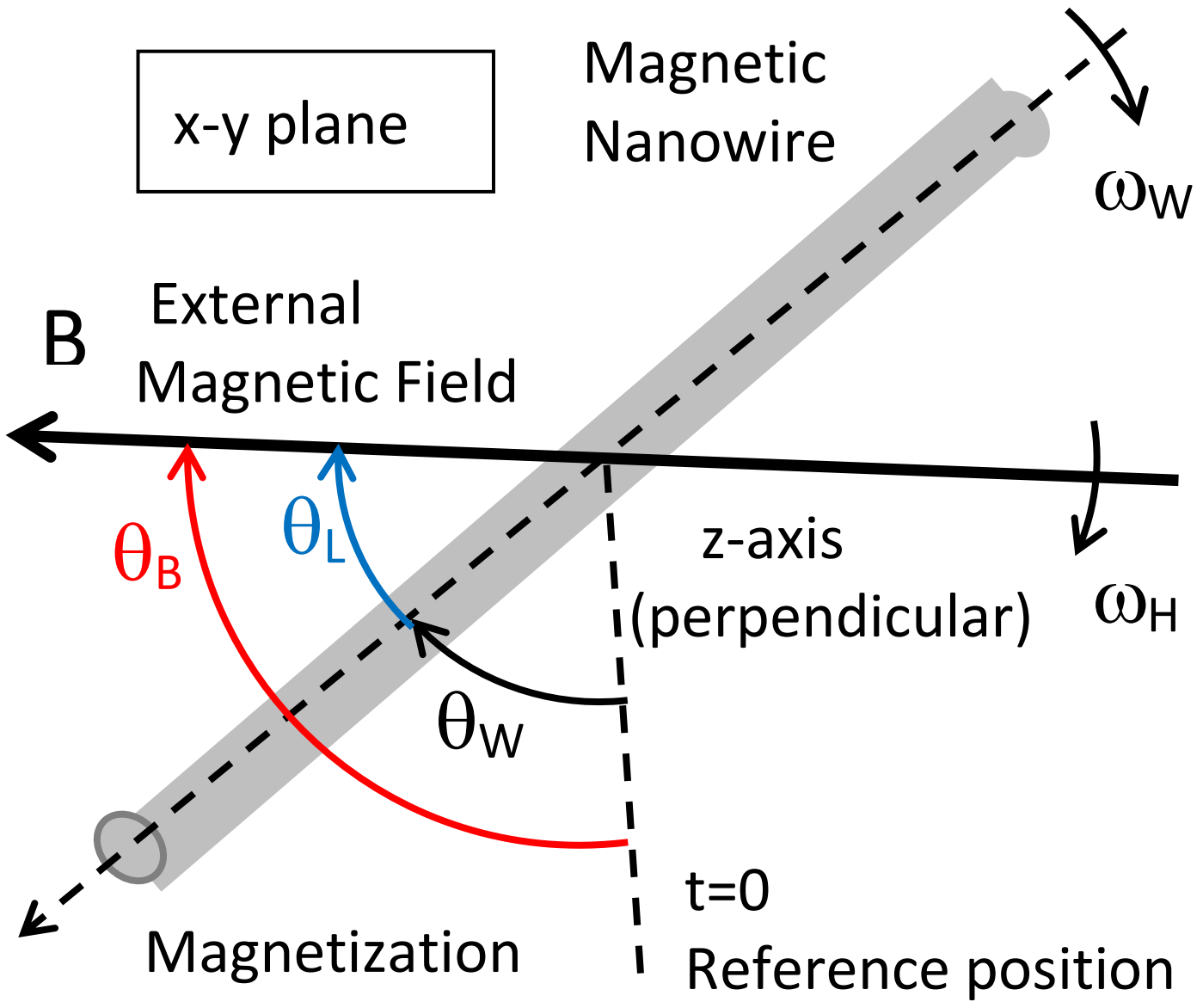

**Figure 6.** Induced rotation of a magnetic nanowire subjected to a rotating magnetic field

An external magnetic field rotating with angular velocity $\omega_H$ in the plane x-y is applied. The premagnetized nanowire tends to follow the rotating field with the long axis laying in the same plane and an instantaneous angle of $\theta_W$. Due to inertia and viscous drag from the surrounding medium, there is a lag angle $\theta_L$ between the nanowire and the field direction, so that $\theta_L = \theta_B - \theta_W = \omega_H t - \theta_W$.

The dynamical equation is

$$T_{Mag} - \alpha \frac{d\theta_w}{dt} = J \frac{d^2\theta_w}{dt^2} \tag{73}$$

The first term in the left-hand side of the above equation is the magnetic torque over the nanowire and the second term represents the viscous torque acting as the resisting mechanism assuming a linear dependence with respect to the angular velocity. J is the inertia moment of the nanowire with respect to the z axis passing through its center. The magnetic torque is given by

$$T_{Mag} = M_s \pi r^2 l B sin\theta_L, \tag{74}$$

where $B$ is the magnetic field density and $M_s$ is the spontaneous magnetization of the nanowire represented as a cylinder of radius r and axial length $l$.

If the term $J\frac{d^2\theta_w}{dt^2}$ may be neglected Eq. (73) may be recast as

$$\frac{d\beta}{dt} = \omega_H - \omega_c sin\beta. \tag{75}$$

In the above equation we have redefined $\theta_L$ as $\beta$ and $\omega_c = M_s\pi r^2 lB/\alpha$.

For reaching a synchronous rotation $sin\beta = \omega_H/\omega_c \leq 1$ implying that $\omega_H \leq \omega_c$. If $\omega_H > \omega_c$ the particle cannot rotate synchronously with the applied field and angle $\beta$ is a periodic function of time. Integration of Eq. (75) leads to an expression similar to Eq. (31)

$$\tan(\beta(t)/2) = D + \sqrt{1-D^2}\ \tan\left(\sqrt{1-D^2}\ \omega_H t/2\right), \tag{76}$$

where $D = \frac{\omega_c}{\omega_H} < 1$ in this case. The period of oscillation of the axis of the nanowire is

$$T = \frac{2\pi/\omega_H}{\sqrt{1-D^2}} \tag{77}$$

If the rotation frequency $\omega_H$ of the external magnetic field changes in a small amount $\Delta\omega_H \ll \omega_H$ coming back to its initial value after a a short interval $\tau$, there will be induced oscillations of the nanowire. In fact, linearizing Eq. (73) with respect to an initial value $\theta_{Ho} = \beta_o$ we obtain that the natural modes of oscillation are of the form $\Delta\beta(t) \sim e^{-(\alpha/2J)t}e^{\pm i\omega_o t}$ with $\omega_0 = \sqrt{4JT_{sinc} - \alpha^2}/2J$ where $T_{sinc} = (M_s\pi r^2 l)cos\beta_0$ is a synchronizing torque which is analogous to the case of a synchronous machine subject to small perturbations [33].

## 5. Analogies with a synchronous reluctance motor

The principle of operation of a typical alternating current machine is based on a rotating magnetic field produced by two or three windings in the stator. As known from undergraduate courses of Electric Machines (ref. [34]), a balanced three phase system of currents of frequency $f$ and phase difference *2π/3* flowing through three windings whose corresponding magnetic axes are disposed with a geometrical angular difference of $2\pi/3$, produces a rotating magnetic field. The same effect occurs for two harmonic currents with the same amplitude and frequency, phase difference $\pi/2$, flowing through two windings whose corresponding magnetic axes are disposed with a geometrical angular difference of $\pi/2$. In the case of a synchronous motor, this rotating magnetic field interacts with the magnetic field produced by a constant current flowing through a winding placed in the rotor of the machine, called the field winding. In steady state conditions a constant electromagnetic torque is developed over the rotor which has two components: one of them corresponds to the alignment torque between the magnetic moment produced by the current flowing through the field winding and the rotating magnetic field, in a form analog to the torque $\vec{p}\times\vec{E}$ acting over a particle with electric dipolar moment not aligned with the applied electric field vector. The other component of the torque arises from the existence of variable reluctance in the air gap of the machine, high reluctance in the quadrature axis and low reluctance in the direct axis [34]. This last component is known as reluctance torque and it is independent of the current through the field winding.

Due to recent developments in power electronic devices, converters and control techniques, reluctance motors have emerged as general and high performance industrial drives for variable speed applications. A reluctance synchronous machine may be described as a synchronous motor with zero field current. Neglecting mechanical losses, in steady state conditions and using Park transformations [35] the electromagnetic torque in its axis is given by [36],

$$T_{em}=\frac{3p}{2}\left(\lambda_d i_q-\lambda_q i_d\right). \tag{78}$$

A brief deduction of this last expression is obtained in Appendix C and considers a coordinate transformation from a system fixed to the stator to other fixed to the rotor, where two axes are identified: one along a magnetic path with minimum reluctance is called direct axis and it is denoted as direct axis ($d$); the other axis along a magnetic path with maximum air gap is called quadrature axis and it is denoted as quadrature axis ($q$).

In Eq. (78) $p$ is the number of pair of poles, $\lambda_d = L_d i_d$ and $\lambda_q = L_q i_q$ are the magnetic flux linkages through the stator windings along $d$ and $q$ axis respectively. $L_d$ and $L_q$, $i_d$ and $i_q$ are the corresponding inductances and currents, respectively. The rotating magnetic field produced by the currents in the stator can be associated to a current vector $\vec{I}$ rotating with angular velocity $\omega = 2\pi f$ with respect to the stator. If at a given instant the angle between the vector $\vec{I}$ and the axis $d$ is $\beta$ the corresponding currents along $d$ and $q$ axes are $i_d = I\cos\beta$ and $i_q = I\sin\beta$ (see appendix A), so that for a two-phase machine with two poles equation (78) may be recast as

$$T_{em} = \left(L_d - L_q\right) i_d i_q = \frac{1}{2} I^2 \left(L_d - L_q\right) \sin 2\beta \,. \tag{79}$$

Comparing the previous equation with Eq.(28) when absorption is negligible, it can be seen that they have a similar form, with the magnitude $I$ of the vector current $\vec{I}$ being analogous to the amplitude of the rotating electric field and with $L_d$ and $L_q$ being analogous to the polarizabilities $\alpha_d$ and $\alpha_q$ of the nanomotor. The reluctance component of electromagnetic torque developed by a conventional three-phase, two-pole synchronous motor can be written as, [34]

$$\omega T_{em} = \frac{1}{2} V^2 \left( \frac{1}{x_q} - \frac{1}{x_d} \right) \sin 2\beta \,, \tag{80}$$

where $V$ is the applied line to line voltage and $x_d$ and $x_q$ are the $d$ and $q$ axes reactances. This last equation is valid if the resistance of the stator winding is neglected and using Park transformation it leads to (79) as shown in Appendix D.

It must be noted that according to the equivalent circuit of a synchronous reluctance motor (SRM) [37-38] the magnitude of the vector current $\vec{I}$ in (79) depends on the applied voltage and also on the velocity. Therefore, an alternative for controlling the torque is acting upon *I* or in the terminal voltage following the variations of the load.

If the SRM drives a load with a torque proportional to the velocity of the form $D\omega_r$ the minimum voltage required for synchronous operation with $\omega_r = \omega$ is determined from a relation similar to that of Eq. (30):

$$\sin 2\beta = \frac{2D\omega^2}{V^2\left(\frac{1}{x_q} - \frac{1}{x_d}\right)} \leq 1 \tag{81}$$

## 6. Numerical examples

For illustrating the conceptual frame presented, we first show two examples where the force and torque are calculated for a system of two dielectric particles. We use a Drude dielectric function with parameters $\varepsilon_b = 9.9 \omega_p = 8.2\ eV$ and $\Gamma = 0.053\ eV$ [13]. Figure 7 shows the average force between a pair of identical gold nanospheres of radii *a* for parallel (solid line) as well as perpendicular (dotted line) excitation. One particle is at the origin, while the other is at z = D. The separation is D = 2.005*a* and we have included multipoles up to order L=40 in the computation. The force is attractive (positive) in the parallel configuration and repulsive (negative) in the perpendicular geometry, as expected. Three multipolar resonances are clearly resolved at this separation, with force peaks of magnitude about $10^4$ larger than the background value. As the separation is increased the resonances move to higher frequencies while fewer become resolved [39]. At a center to center separation of about three particle radii and larger, only one resonance is seen. This dipolar peak has been included in the figure for comparison, with an amplification factor of 5000 (dashed curve).

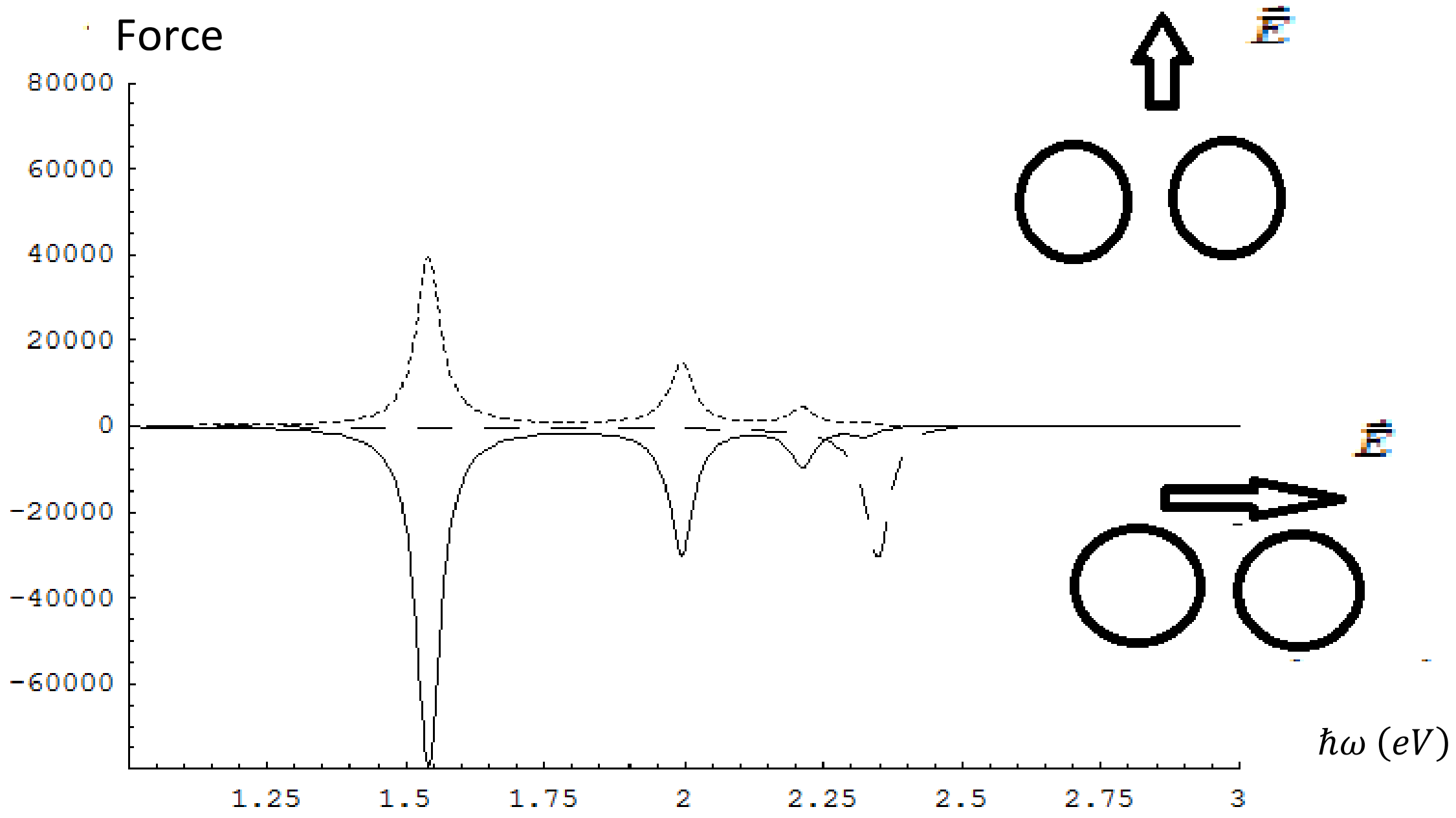

**Figure 7.** Electric force between two identical spheres as a function of frequency for two gold nanospheres, with separation 2.005$a$ between their centers. Solid and dotted curves correspond to parallel and perpendicular excitation, respectively, calculated with L=40. Dashed curve corresponds to the average force calculated for parallel excitation and a separation 3$a$, with an amplification factor 5000. The angular frequency of oscillation of the applied field is expressed in $eV$ equivalent and the force in units of $E_0^2 a^2$ where $E_0$ is the amplitude of the applied electric field.

In Fig. 8 we show the $z$ component of the average torque acting on each nanoparticle as given by Eq. (72). Separations are 2.005$a$ (solid curve) and 3$a$ (dashed curve). The pair is subject to an electric field whose direction rotates in the plane $xy$. As for the force, several resonances are resolved at small separation, while beyond about separation $3a$ only one peak is observed. It can be seen that as the spheres become closer other resonances occur for frequencies below the value $\omega = \omega_p / \sqrt{\varepsilon_b - 1}$ corresponding to that obtained in the dipolar approach. These additional resonance frequencies correspond to resonant modes associated to the multipole moments.

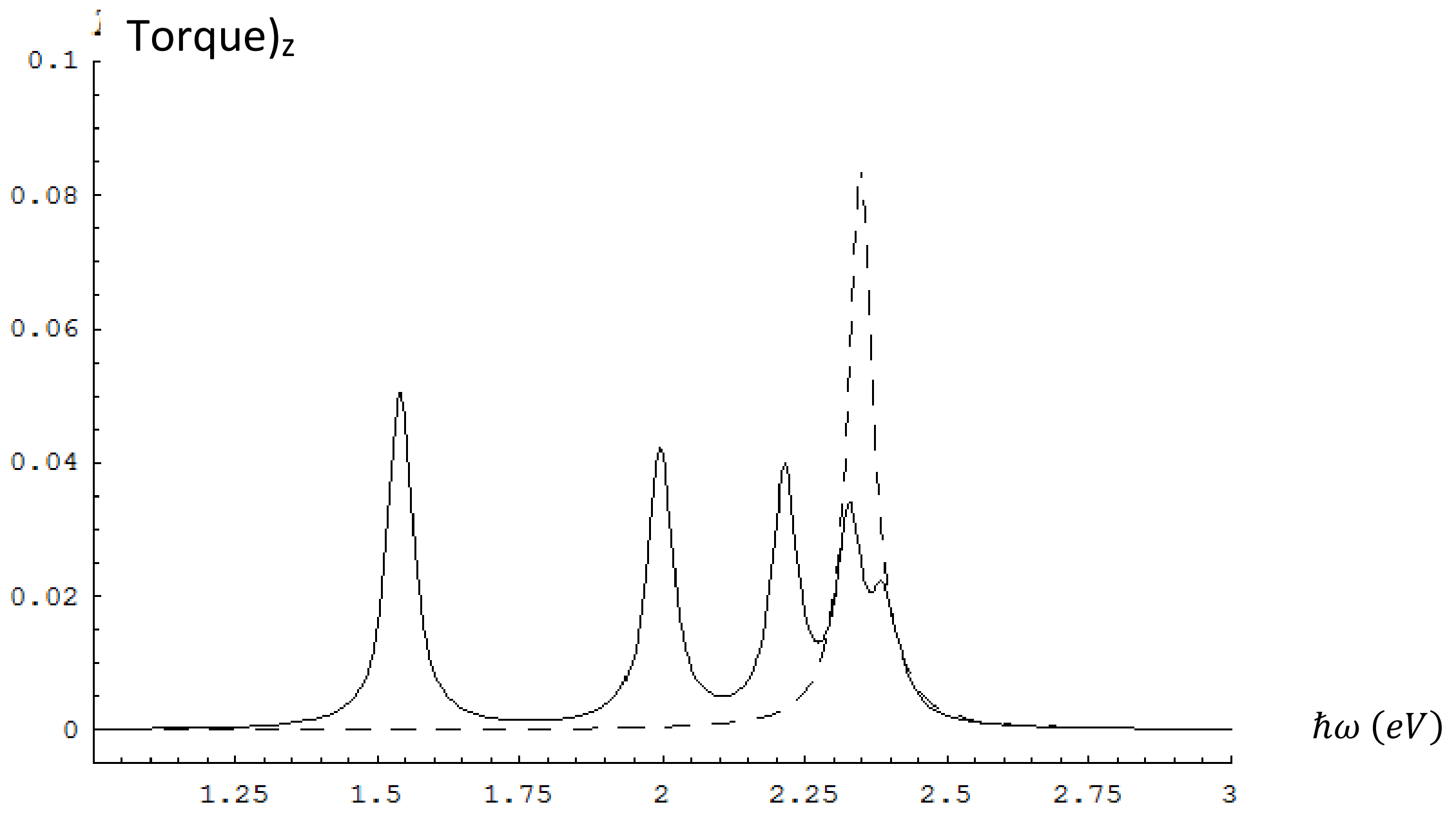

**Figure 8.** Time average torque over each sphere of a pair gold nanospheres subjected to a rotating electric field, as a function of the frequency, for separation between centers 2.005*a* (solid curve) and 3*a* (dashed curve), where *a* is the radii of each sphere. Results shown are for L=40 and L=10, respectively. The angular frequency ω is translated to eV and the torque is in units of $E_0^2 a^2$ where $E_0$ is the amplitude of the applied electric field.

Now we show two examples where the rotation of a synchronous nanomotor (SN) and a synchronous reluctance motor (SRM) is studied. For the case of the SN we consider a nanoparticle with the same parameters used in [16]. For the synchronous reluctance motor the parameters considered are: inertia moment of the rotor $J$ = 0.01 kg m$^2$, damping coefficient $\gamma$= 0.1 J / rad, $L_d$ = 0.15 pu, $L_q$= 0.08 pu, rated power 1 kVA, rated three phase voltage 230 V. In the first example, we study the starting process of both devices for a frequency of 5 Hz of the applied field. In Fig. 9 solid (dashed) blue curve corresponds to the starting of the nanoparticle with amplitude of the electric field twice (equal) the minimum value $2\times10^{7}$ V/m, required for reaching the synchronous rotation velocity of 5 rev/s. Solid (dashed) red curve correspond to the starting of the synchronous reluctance motor with voltage twice (equal) the minimum value required for reaching the synchronization. This value is equal to 20 % of the nominal value.

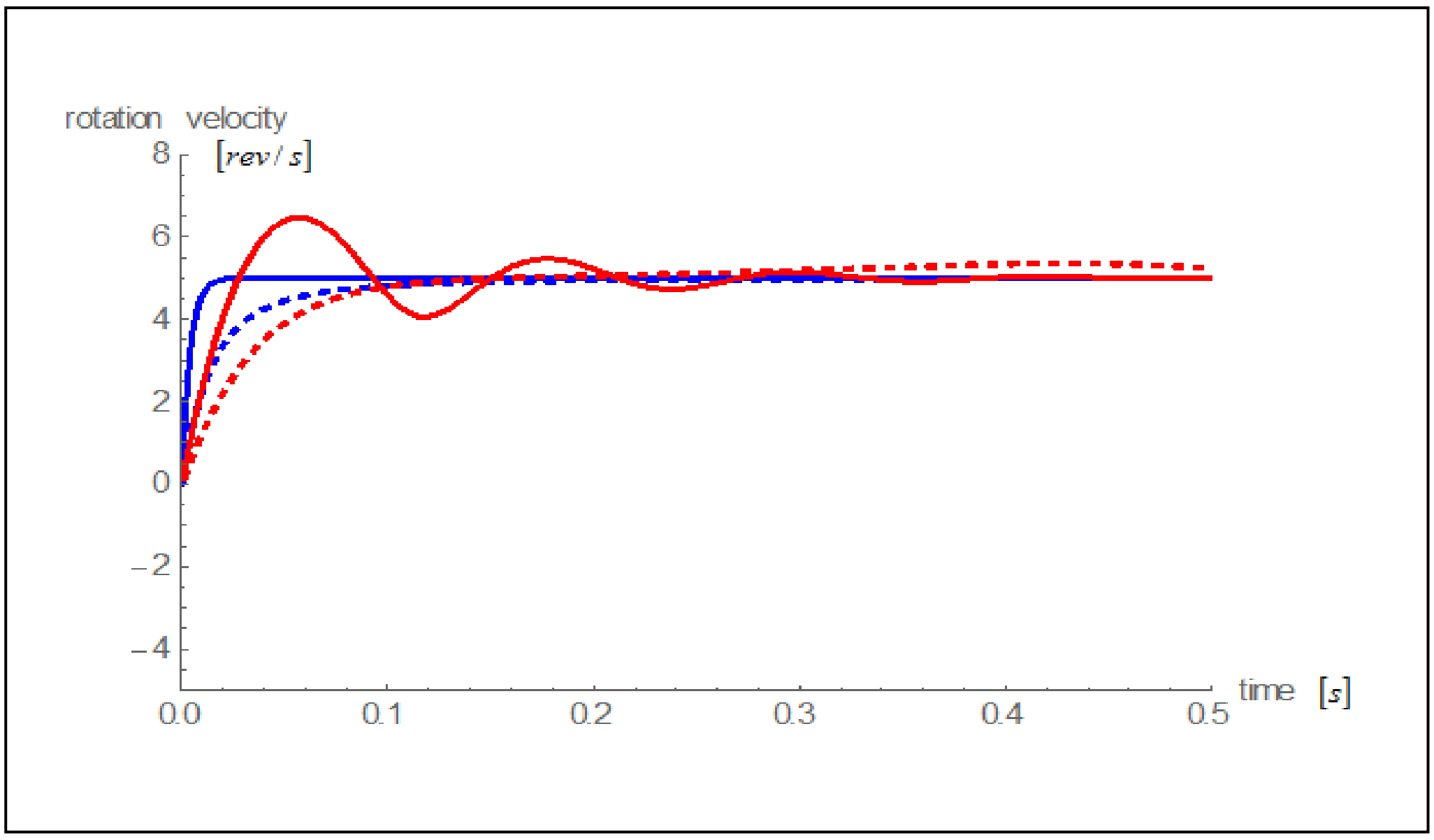

**Figure 9.** Starting process of a SN and a SRM. Solid (dashed) blue curve corresponds to the rotation velocity of the nanomotor in rev/s, with amplitude of the electric field twice (equal) the minimum value $2\times10^{7}$ V/m required for reaching the synchronous rotation velocity of 5 rev/s. Solid (dashed) red curve correspond to the velocity in rev/s of the SRM with a voltage twice (equal) the minimum value necessary for reaching the synchronous condition.

As can be seen in Fig. 9 for the case of the synchronous nanomotor as the amplitude of the electric field increases the lower is the time required for reaching the synchronous velocity. The same result is valid for the SRM except that in this case with an applied voltage twice the minimum value for synchronization, there is an oscillatory behavior before reaching the synchronous velocity. This behavior can be explained considering that as previously noted the magnitude of the vector current $\vec{I}$ in Eq. (79) and therefore the amplitude of the rotating magnetic field depends on the velocity, while in the case of the synchronous nanomotor the electric torque depends only on the electric field imposed by the source.

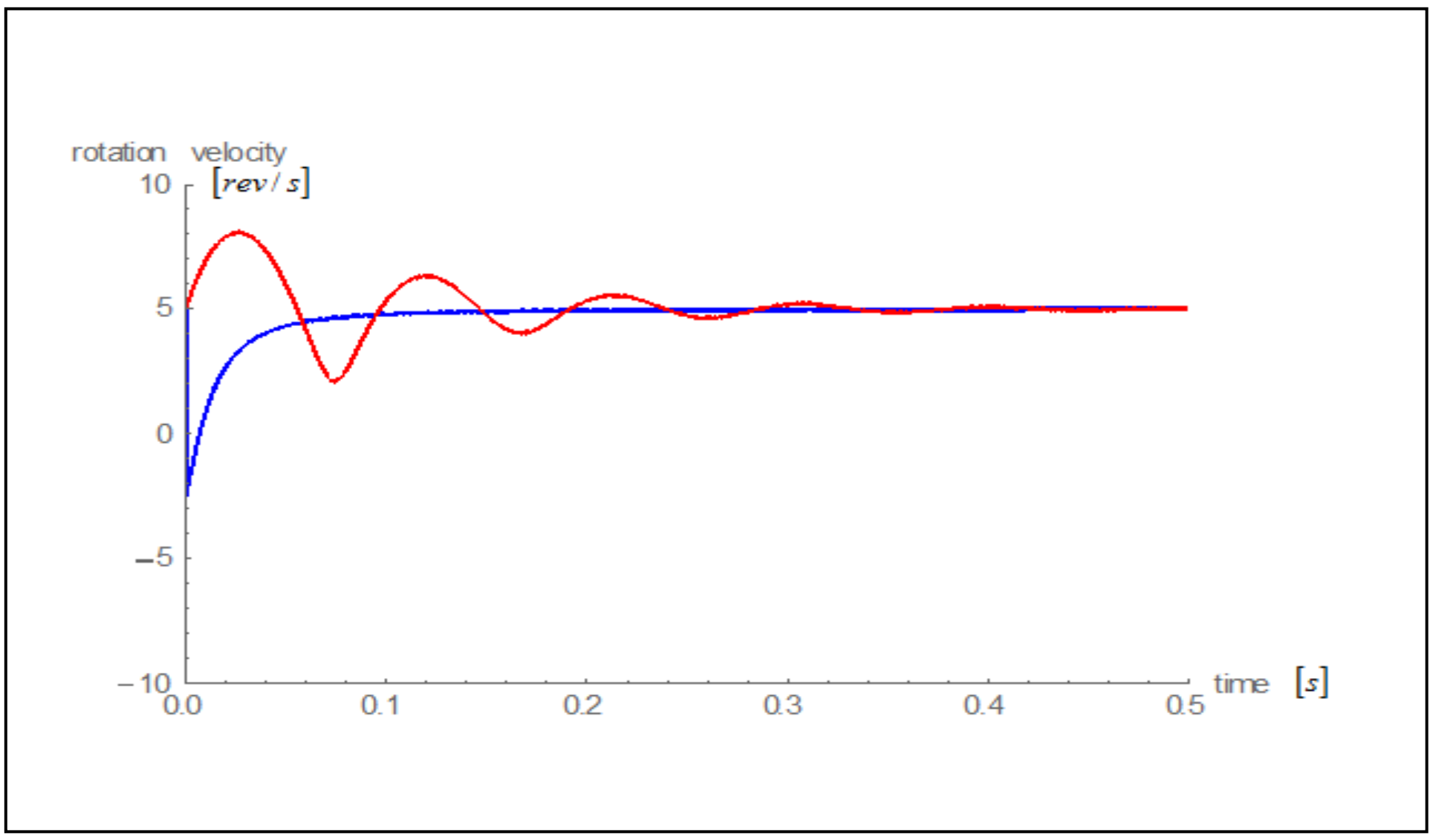

**Figure 10.** Transient response of the SN (blue curve) and the SRM (red curve) under a sudden decrease of 50 % in the amplitude of the electric field or in the applied voltage, respectively.

As the second example, in Fig. 10 we show the transient response of the SN (blue curve) and the synchronous reluctance motor (red curve) under a sudden decrease of 50 % in the amplitude of the electric field or in the applied voltage, respectively. Initially these devices are rotating at synchronous velocities of 5 rev/s with a rotating electric field of amplitude $4\times10^7$ V/m for the SN and with the nominal voltage for the SRM. Results are shown in Fig. 10 and it can be seen that in both cases after a transient process the velocity returns to the synchronous value, with oscillations in the case of the SRM as discussed in relation with Fig. 9.

These examples show that the rotation velocity of the synchronous nanomotor and of the SRM can be controlled by the frequency of the applied electric field or of the voltage, respectively. In the starting process of both devices the velocity reaches the synchronous velocity corresponding to the frequency of the rotating field, provided the amplitude of the electric field or the applied voltage, respectively, is not lower than minimum values determined from (30) and (81).

In the case of the starting of the SRM there are oscillations before reaching the synchronous velocity which can be eliminated by controlling the current, as discussed in [38]. Under a sudden reduction in the amplitude of the electric field or in the voltage, after a transient process the velocity can return to the synchronous value provided that these amplitudes are not lower than the corresponding minimum values.

## 7. Conclusions

A conceptual frame has been presented showing how basic concepts acquired in courses of Classical Electromagnetism and Energy Conversion of an undergraduate program in electric engineering, may be integrated for learning how it is possible to induce by means of circularly polarized light a synchronous rotation of a nanoparticle. The analogy with the operation of a synchronous reluctance motor (SRM) has been studied. The common principle is the transfer of angular momentum from a rotating electromagnetic field to a dielectric or magnetic system.

Starting from basic concepts of electromagnetism and using the concepts of polarizability and dielectric functions, our work analyzes the transfer of angular momentum from the electromagnetic field to isotropic and non-isotropic objects studying in particular the induced motion of an object with non-homogeneous polarizability. This is the case of a birrefringent nanoscopic element subject to a rotating electric field, where we study conditions for synchronous rotation so that it operates as a synchronous nanomotor.

By means of numerical examples, it has been shown that in steady state conditions the velocity of rotation of the synchronous nanomotor and the SRM depends on the frequency of the electric field or of the applied voltage, respectively. The control of the torque of the synchronous nanomotor or the synchronous reluctance motor, can be achieved by acting upon the amplitude of the rotating electric field or the injected current, respectively.

## ACKNOWLEDGMENT

The authors thank to Escuela de Ingeniería Eléctrica of Pontificia Universidad Católica de Valparaíso and to the Physics Department of Universidad Santa María, for the support given during the elaboration of this review.

## Appendix A. Lorentz model for obtaining the dielectric function

This appendix contains a derivation of Eqs. (1) to (4) according to the Lorentz model for dielectrics, in which each atom subject to an oscilating electric field $\vec{E} = \vec{E}_0 e^{-i\omega t}$ is represented as a mechanical oscillator with elastic constant $k$ and damping $\gamma$. The driving force is due to the electric force acting over each electron (mass $m$) so that the corresponding dynamics equation is

$$m\frac{d^2\vec{r}}{dt^2} + \gamma\frac{d\vec{r}}{dt} + k\vec{r} = e\vec{E} \qquad \text{(A-1)}$$

Considering a solution of the form $\vec{r} = \vec{r}_0 e^{-i\omega t}$ and inserting in Eq. (A-1) we obtain

$$\vec{r}_0 = \frac{e}{k - m\omega^2 - i\gamma\omega}\,\vec{E}_0. \qquad \text{(A-2)}$$

With $\omega_{oe}^2 \equiv k/m$ and $\Gamma_d \equiv \gamma/m$ the above equation may be recast as

$$\vec{r}_0 = \frac{e/m}{\omega_{oe}^2 - \omega^2 - i\Gamma_d\omega}\vec{E}_0 \qquad \text{(A-3)}$$

For an atom with Z electrons, the corresponding dipolar moment is $\vec{p} = Ze\vec{r} = \alpha\vec{E}$ where $\alpha$ is the atomic polarizability and therefore the above equations lead to

$$\alpha(\omega) = \frac{Ze^2/m}{\omega_{oe}^2 - \omega^2 - i\Gamma_d\omega} \qquad \text{(A-4)}$$

If $n$ is the concentration of atoms, each one with Z electrons, the polarization vector is

$$\vec{P} = n\alpha\vec{E} = \frac{nZe^2/m}{\omega_{oe}^2 - \omega^2 - i\Gamma_d\omega}\vec{E} \qquad \text{(A-5)}$$

The displacement vector $\vec{D}$ is related with the electric field $\vec{E}$ as

$$\vec{D} = \epsilon_o\vec{E} + \vec{P} = \epsilon_o\vec{E} + n\alpha\vec{E} = \epsilon(\omega)\epsilon_o\vec{E} \qquad \text{(A-6)}$$

Therefore the dielectric function $\epsilon(\omega)$ is given by

$$\epsilon(\omega) = 1 + \frac{n\alpha}{\epsilon_o} = 1 + \frac{nZe^2/m\epsilon_o}{\omega_{oe}^2 - \omega^2 - i\Gamma_d\omega} = 1 + \frac{\omega_{pe}^2}{\omega_{oe}^2 - \omega^2 - i\Gamma_d\omega} \qquad \text{(A-7)}$$

where $\omega_{pe}$ is the plasma frequency given by $\omega_{pe}^2 = nZe^2/m\epsilon_0$.

## Appendix B. Forces between electric dipoles in the dipolar approach

In this appendix we study the interaction of two spherical particles in the dipole approximation in order to get equations (6) and (7). From classical electromagnetism it is known that the electric potential of a point dipole $\vec{p}$ placed at the origin, in a point placed at position $\vec{r}$ as shown in Fig. B1 is given by

$$V(\vec{r}) = \frac{\vec{p} \cdot \hat{r}}{4\pi\varepsilon_o r^2} = \frac{p cos\theta}{4\pi\varepsilon_o r^2} \qquad \text{(B-1)}$$

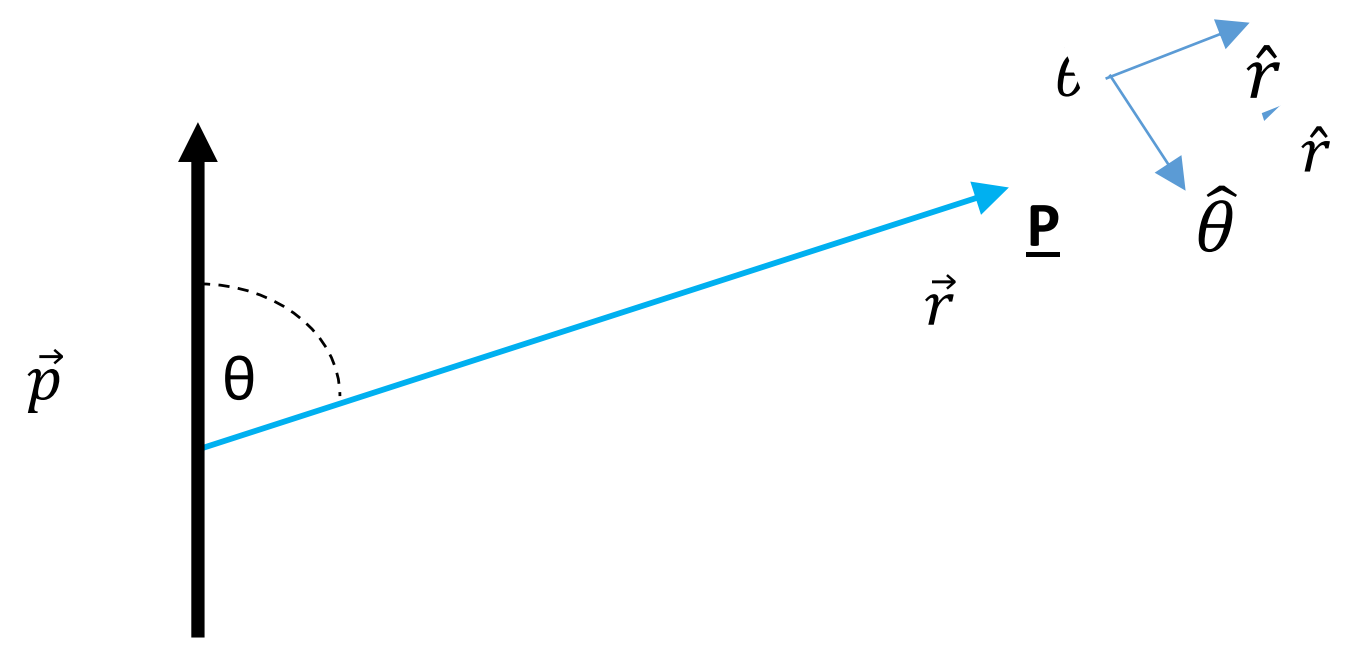

Figure B1. Electric dipole.

The corresponding electric field is obtained from $\vec{E} = -\nabla V$ resulting in the following expression in spherical coordinates:

$$\vec{E}(\vec{r}) = \frac{p}{4\pi\varepsilon_o} \frac{1}{r^3} \left(\hat{r}\ 2cos\theta + \hat{\theta}\ sin\theta\right) \qquad \text{(B-2)}$$

Where $\hat{r}$ and $\hat{\theta}$ are unitary vectors in the radial and tangential directions, respectively, as shown in Fig. B1.

Let us consider now another electric dipole $\vec{p}_2 = |q|\ \vec{a}$ with center placed at point **P** so that the corresponding negative and positive charges, are at positions $\vec{r}_-$ and $\vec{r}_+$ with respect to the origin at center of the dipole $\vec{p}$, so that $\vec{r}_\pm = \vec{r} \pm \vec{a}/2$. A series expansion of the electric field produced by $\vec{p}$ gives the following expression for the electric field acting over the individual charges $+q$ and $-q$ of dipole $\vec{p}_2$:

$$\vec{E}(\vec{r}_{\pm}) = \vec{E}(\vec{r}) \pm \left(\frac{1}{2}\vec{a}\cdot\nabla\right)\vec{E}\Big|_{\vec{r}} + \left(\frac{1}{2}\vec{a}\cdot\nabla\right)^2\vec{E}\Big|_{\vec{r}} \pm \frac{1}{6}\left(\frac{1}{2}\vec{a}\cdot\nabla\right)^3\vec{E}\Big|_{\vec{r}} + \cdots \quad \text{(B-3)}$$

Therefore the net force over dipole $\vec{p}_2$ is:

$$\vec{F} = |q|\{\vec{E}(\vec{r}_+) - \vec{E}(\vec{r}_-)\} = (\vec{p}_2\cdot\nabla)\vec{E}\Big|_{\vec{r}} + \frac{1}{24}(\vec{p}_2\cdot\nabla)^3\vec{E}\Big|_{\vec{r}} + \cdots \quad \text{(B-4)}$$

If $|\vec{a}| \ll |\vec{r}|$ is valid the approach of keeping only the first term in the above expression. In spherical coordinates this leads to the following expressions for the radial and tangential components of the electric force between two electric dipoles:

$$F_r = -3k\left\{\frac{2p_{1r}p_{2r}}{r^4} - \frac{p_{1\theta}p_{2\theta}}{r^4}\right\}, \quad \text{(B-5)}$$

$$F_\theta = -\frac{6kp_{1\theta}p_{2r}}{r^4}, \quad \text{(B-6)}$$

where $p_{1r}$, $p_{2r}$ and $p_{1\theta}$, $p_{2\theta}$ are the radial and tangential components of the electric dipoles $\vec{p} = \vec{p}_1$ and $\vec{p}_2$ respectively. Because $\vec{p}_1$ is at the origin the radial direction is along $\vec{r}$ that points to the center of $\vec{p}_2$ and is at angle $\theta$ with $\vec{p}_1$ because we choose the z-axis along $\vec{p}_1$. The constant $k$ depends on the system of units and in Gaussian units is 1. Then we assume that the system of two particles is subjected to an oscillating electric field along the z-axis: $\vec{E}(t) = \hat{z}E_o e^{-i\omega t}$ .

The radial and tangential components of the electric dipole moments induced in each sphere under the action of this oscillating electric field are:

$$p_{1r} = p_{2r} = \alpha_s E_o cos\theta, \quad \text{(B-7)}$$

$$p_{1\theta} = p_{2\theta} = \alpha_s E_o sin\theta, \quad \text{(B-8)}$$

where $\alpha_s$ is the polarizability of a sphere of radius $a$, which in Gausssian units is given by [22]

$$\alpha_s = a^3[\epsilon(\omega) + 1]/[\epsilon(\omega) - 2], \quad \text{(B-9)}$$

where $\epsilon(\omega)$ is the dielectric function of the sphere. In general the dielectric function depends on the frequency and is a complex quantity whose imaginary part represents absorption or dielectric losses.

Following Eqs. (B-7) and (B-8) we define effective polarizabilities $\alpha_r$ and $\alpha_\theta$ of spheres which take into account the electromagnetic coupling between their dipole moments. Then,

$$p_r = \alpha_s \left\{E_o cos\theta + \frac{2p_r}{r^3}\right\} = \alpha_r E_o cos\theta \qquad \text{(B-10)}$$

$$p_\theta = \alpha_s \left\{E_o sin\theta - \frac{p_\theta}{r^3}\right\} = \alpha_\theta E_o sin\theta. \qquad \text{(B-11)}$$

From Eqs. (B-10) and (B-11) the effective polarizabilities are

$$\alpha_r = \frac{a^3/3}{n_A - u}, \qquad \text{(B-12)}$$

$$\alpha_\theta = \frac{a^3/3}{n_R - u}, \qquad \text{(B-13)}$$

where $a$ is the radius of spheres, $n_A = \frac{1}{3}(1 - 1/4\sigma^3)$, $n_R = \frac{1}{3}(1 + 1/8\sigma^3)$, $\sigma = r/2a$ and $u = 1/[1 - \epsilon(\omega)]$ is the complex spectral variable.

Combining the above equations and taking time-averaged values for the components of the force leads to equations (B-14) and (B-15):

$$F_r(\vec{r}) = \frac{a^2 E_o^2}{96\sigma^4}\left[\frac{\sin^2\theta}{|n_R - u|^2} - \frac{2\cos^2\theta}{|n_A - u|^2}\right] \qquad \text{(B-14)}$$

$$F_\theta(\vec{r}) = \frac{-a^2 E_o^2}{96\sigma^4}\operatorname{Re}\left[\frac{\sin 2\theta}{(n_A - u)(n_R - u)^*}\right] \qquad \text{(B-15)}$$

# Appendix C. Electric force between a pair of interacting particles considering higher order multipoles

In this appendix it is explained how the model described in section 3.4 can be used for determining the induced force between two dielectric spheres of radius $a$ subject to an external electric field. In particular and as examples of application of this formalism we show the calculation of the tangential and radial componenst of this force.
This section is based in ref. [28]. All the expressions used here are in Gaussian units.
We determine the limit of Eqs. (64) and (65) when the distance between the centers of two interacting dielectric spheres is greater than three times their radii so that higher order induced multipoles may be neglected and the dipolar approach is acceptable. For that purpose as shown in Fig. C1 we consider two identical spheres placed at positions such that an applied electric field has a direction forming an angle δ with respect to a line between their centers.

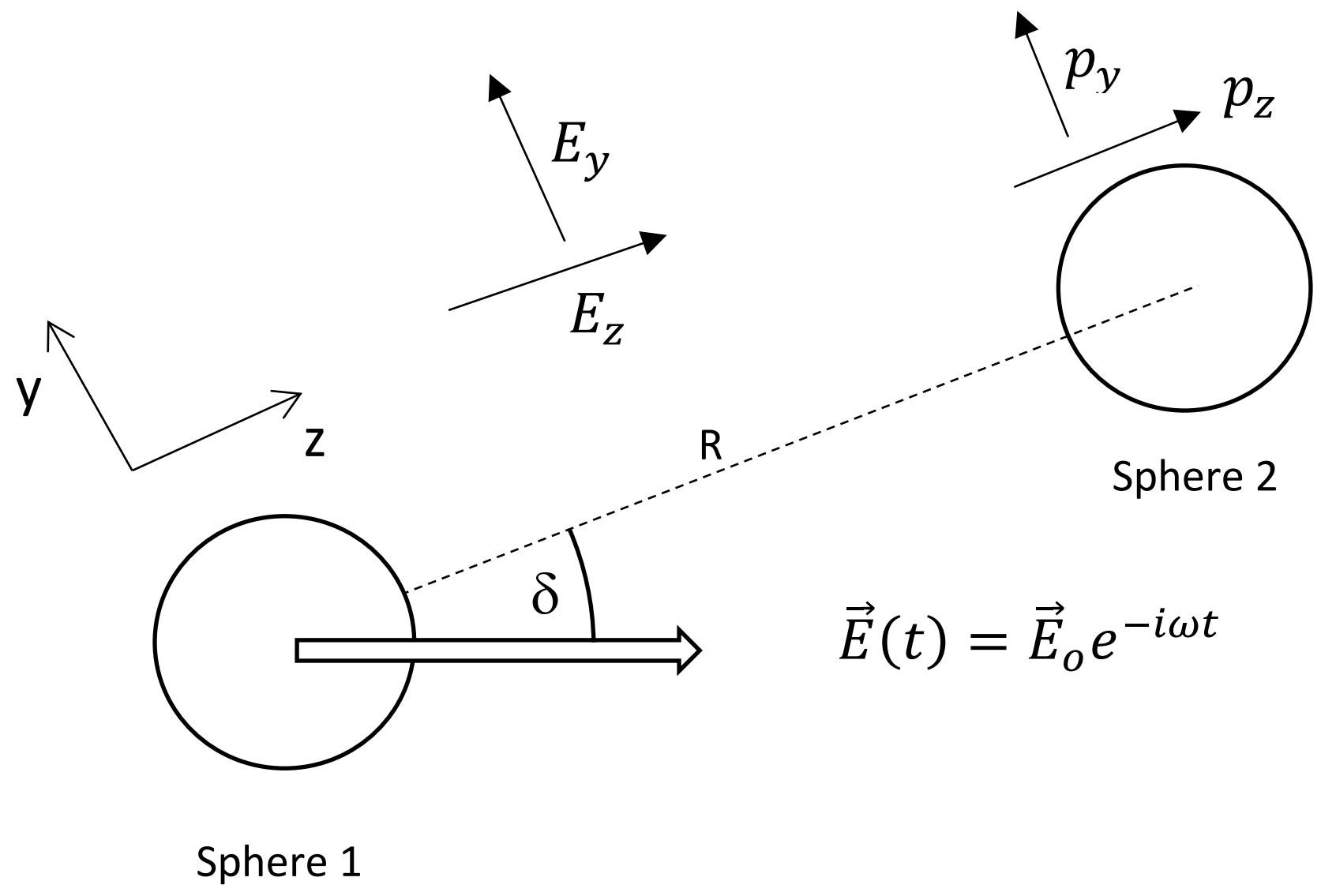

Figure C1. Two spheres placed so that an applied electrical field $\vec{E}$(t) is at an angle $\delta$ with respect to a line between their centers.

First, we calculate the tangential component of this force which according to Fig. C1 corresponds to the y direction. Starting from Eq. (65) for the case of a pair of spheres subjected to an electric field with components along directions z and *y*, the component of the force in the *y* direction may be written as

$$\left\langle F_{iy}\right\rangle = -\pi \operatorname{Re} i \sum_{l=1}^{\infty}\left(\alpha_{l+1,i}\right)^{-1}\left[2\sqrt{\frac{(l+1)l}{(2l+1)(2l+3)}}q_{l+1,0,i}q_{l,1,i}^{*} + 2\sqrt{\frac{(l+2)(l+1)}{(2l+1)(2l-1)}}q_{l+1,1,i}q_{l,0,i}^{*}\right] \quad \text{(C-1)}$$

In this last equation it has been considered that in this case for active modes the possible values of *m* are: *m=0* and *m= ±1* for the parallel and perpendicular component of the electric field, respectively.

According to Eq. (5) of Reference [27]:

$$q_{l+1,m,i} = -\frac{2(l+1)+1}{4\pi}\alpha_{l+1,i}b_{l+1,m,i}, \quad \text{(C-2)}$$

where $b_{l+1,m,i} = \sum_{l'=1}^{\infty}(-1)^{l'} A_{l+1,m,i}^{l',m,j}\, q_{l',m,j}$ for *m*= 0, ±1. In this case indexes *i* or *j* identify the corresponding sphere (1 or 2) for which the tangential force is calculated and it is considered that according to Eq. (19) of ref. [27] for the case of two identical spheres there is coupling only between modes with the same value of *m*. Furthermore since we are considering the dipolar approach the term $V_{lm}^{ext}(i)$ is zero for $l > 1$. The coupling coefficient $A_{l+1,m,i}^{l',m,j}$ for this case is given by

$$A_{l+1,m,i}^{l'mj} = 4\pi\binom{l+l'+1}{l+1}\frac{1}{\left[(2l+3)(2l'+1)\right]^{1/2}}\frac{1}{R^{l+l'+2}} \quad \text{if m=0,} \quad \text{(C-3)}$$

$$A_{l+1,m,i}^{l'mj} = -\sqrt{\frac{(l+1)l'}{(l+2)(l'+1)}}A_{l+1,0,i}^{l'0j} \quad \text{if m=±1.} \quad \text{(C-4)}$$

Considering the definition of the normalized multipole moments $\bar{q}_{l,m,i} = q_{l,m,i}/a^{l}$, the relation $q_{l',m,j} = (-1)^{l'+1}\bar{q}_{l',m,i}$ between induced multipoles of order *l',m* in each sphere, replacing the above expressions in Eq. (C-1), and after some algebra, we obtain the following expression for the time-average value of the tangential force

$$\left\langle F_{iy}\right\rangle = -\frac{-2\pi}{a^{2}}\sin\delta\cos\delta \operatorname{Re} i\sum_{l=1}^{\infty}\sum_{l'=1}^{\infty}\left\{ f(l,l')\bar{s}_{l'0i}\ \bar{s}_{l,1,i}^{*}\ F(l,l')\bar{s}_{l',1,i}\ \bar{s}_{l,0,i}^{*}\right\}\left(\frac{a}{R}\right)^{l+l'+2} \quad \text{(C-5)}$$

where

$$f(l,l')=\frac{(l+l'+1)!}{(l+1)!l'!}\frac{1}{[(2l+1)(2l'+1)]^{1/2}}[l(l+1)]^{1/2} \quad \text{(C-6)}$$

$$F(l,l')=-\frac{(l+l'+1)!}{(l+1)!l'!}\frac{1}{[(2l+1)(2l'+1)]^{1/2}}\left[\frac{(l+1)l'}{(l+2)(l'+1)}\right]^{1/2}[(l+2)(l+1)]^{1/2} \quad \text{(C-7)}$$

In Eq. (C-5) we have denoted as $\bar{s}_{l,0,i}$ and $\bar{s}_{l,\pm 1,i}$ the normalized multipole moments calculated for $\delta = 0$ and $\delta = \pi/2$, respectively, corresponding to only parallel (perpendicular) excitation, so that the electric field vector is along the z ($y$) direction in Fig. C1.

In the limit where the dipolar approach is valid and higher order multipoles may be neglected, from the above equations it can be shown that the resulting tangential force has a time-average given by $\langle F_{e\theta}\rangle = -(6/R^4)\frac{1}{2}Re\left[p_z p_y^*\right]$. This expression coincides with Eq. (2) of ref. [13]. In obtaining this limit we have considered relations $a\bar{s}_{1,0,1} = \sqrt{3/4\pi}\,p_z$ and $a\bar{s}_{1,\pm 1,1} = i\sqrt{3/4\pi}\,p_y$ [22].

As a second example we calculate the force for the case δ=0. From Eqs. (66) and (71) and considering that in this case only modes with m=0 are excited, we obtain

$$\langle F_{iz}\rangle = 2\pi\, Re\sum_l \frac{(l+1)(\varepsilon+1)+1}{(l+1)(\varepsilon-1)a_i^{2l+3}}\sqrt{\frac{(l+1)(l+1)}{(2l+3)(2l+1)}}\,q_{l+1,0,i}q_{l,0,i}^* \quad \text{(C-8)}$$

As the separation between the centers of the spheres increases, the dipolar approach becomes valid so for R >> a, only the term with l=1 is kept and after a development similar to the above case we get a result coinciding with Eq. (1) of ref. [13] for the case of parallel excitation:

$$\langle F_{iz}\rangle = Re[3p_z p_z^*/R^4]. \quad \text{(C-9)}$$

# Appendix D. Park Transformation

Park's transformation is a phase transformation (coordinate transformation) between the three physical phases in a three phase system and three new phases, or coordinates, which are convenient for the transient analysis of synchronous machines under electromechanical oscillations produced by perturbations. This transformation is also known as the dq–transformation.

The phase quantities in the a–, b–, and c– phases will vary periodically in steady state. Further, the self and mutual inductances between stator circuits and rotor circuits will vary with the rotor position.

Instead of performing all computations in the fixed stator system, the stator quantities voltages, currents, and fluxes can be transformed to a system that rotates with the rotor. Thus, two orthogonal axes are defined as shown in Fig. D1. One is along the axis in which the current in the rotor windings generates a flux, and other in an axis perpendicular to this. The first is the direct axis (d–axis), and the other is the quadrature axis (q–axis). From now on, the denominations d–axis and q–axis will be used. To make the system complete, a third component corresponding to the zero sequence must be considered.

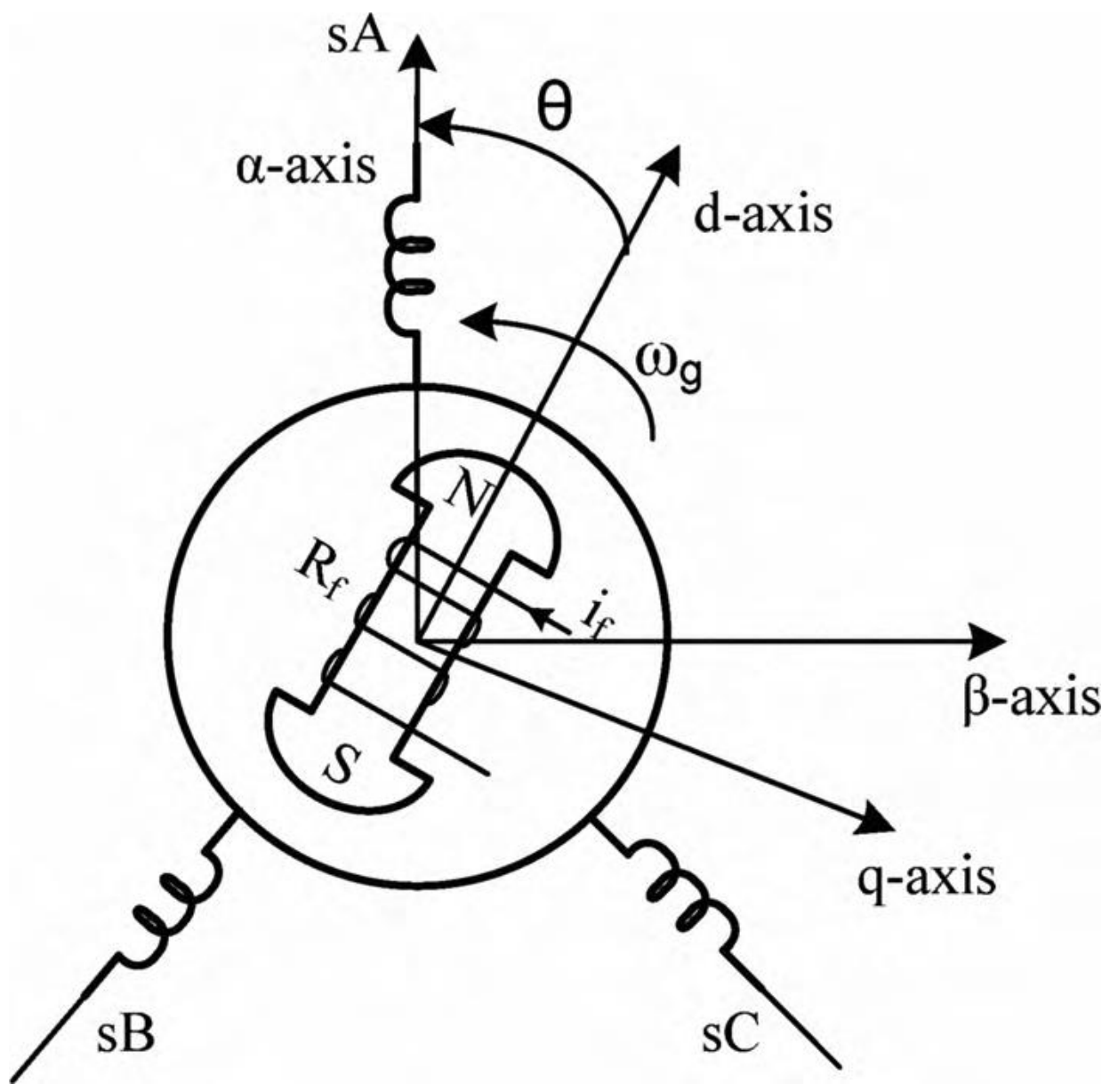

Figure D1. Rotating d-axis and q-axis, and fixed α-axis and β-axis.

The machine in Fig. D1 has one pole pair, but Park's transformation can, of course, be applied to machines with an arbitrary number of pole pairs.

According to Park's transformation, the connection between the phase currents and the transformed currents is given by

$$i_o = \sqrt{\frac{1}{3}}(i_a + i_b + i_c) \tag{D-1}$$

$$i_d = \sqrt{\frac{2}{3}}\left(i_a cos\theta + i_b cos\left(\theta - \frac{2\pi}{3}\right) + i_c cos\left(\theta + \frac{2\pi}{3}\right)\right) \tag{D-2}$$

$$i_q = \sqrt{\frac{2}{3}}\left(-i_a sin\theta - i_b sin\left(\theta - \frac{2\pi}{3}\right) - i_c sin\left(\theta + \frac{2\pi}{3}\right)\right) \tag{D-3}$$

If the *a*–axis is chosen as reference,

$$\theta = \omega_g t + \theta_o \ , \tag{D-4}$$

It should be pointed out that $i_a$, $i_b$ and $i_c$ are the real physical phase currents as functions of time and not a phasor representation of those. Now, we can define

$$x_{abc} = (x_a, x_b, x_c)^T, \tag{D-5}$$

$$x_{0dq} = \left(x_0, x_d, x_q\right)^T, \tag{D-6}$$

where *x* can be an arbitrary quantity, like voltage, current, or flux. With this notation, Park's transformation can be written as the following matricial relation:

$$x_{0dq} = P x_{abc} \tag{D-7}$$

Where the matrix P is

$$P = \sqrt{\frac{2}{3}}\begin{bmatrix} 1/\sqrt{2} & 1/\sqrt{2} & 1/\sqrt{2} \\ cos\theta & cos(\theta - 2\pi/3) & cos(\theta + 2\pi/3) \\ -sin\theta & -sin(\theta - 2\pi/3) & -sin(\theta + 2\pi/3) \end{bmatrix}. \tag{D-8}$$

It can be shown that $P^{-1} = P^T$ so that Park's transformation is orthonormal. This implies that this transformation is power invariant:

$$p = u_a i_a + u_b i_b + u_c i_c = u_d i_d + u_q i_q + u_o i_o \tag{D-9}$$

For a three-phase balanced system of angular frequency ω the phase currents in steady-state are

$$i_a(t) = \sqrt{2}Icos\omega t\ , \qquad \text{(D-10a)}$$
$$i_b(t) = \sqrt{2}Icos(\omega t - 2\pi/3), \qquad \text{(D-10b)}$$
$$i_c(t) = \sqrt{2}Icos(\omega t + 2\pi/3), \qquad \text{(D-10c)}$$

where $I$ is the rms current.

From Eqs. (D-2), (D-3) and (D-4) one obtains that if the rotor runs synchronously with the rotating magnetic field produced by the currents in the stator $(\omega_g = \omega)$,

$$i_d = I\sqrt{3}\ cos\theta, \qquad \text{(D-11a)}$$
$$i_q = I\sqrt{3}\ sin\theta. \qquad \text{(D-11b)}$$

As shown in ref. [32] the voltage equations in the d-q system for a balanced system are

$$u_d = -ri_d - L_d\frac{di_d}{dt} + \omega L_q i_q \qquad \text{(D-12a)}$$

$$u_q = -ri_q - L_q\frac{di_q}{dt} - \omega L_d i_d \qquad \text{(D-12b)}$$

In the previous equations *r* corresponds to the stator phase resistances, $L_d$ and $L_q$ are the inductances in *d* and *q* axis, respectively.

Equations (D-9) and (D-12) and a power balance lead to

$$p = u_d i_d + u_q i_q = \omega\left(L_q - L_d\right)i_d i_q + \frac{d}{dt}\left(\frac{L_d}{2}i_d^2 + \frac{L_q}{2}i_q^2\right) \qquad \text{(D-13)}$$

This last equation is valid if the resistance *r* can be neglected and the second term represents the time variation of the energy stored in the electromagnetic field while the first term $\omega\left(L_q - L_d\right)i_d i_q$ is the mechanical power for synchronous operation. This expression is equivalent to that learned in courses of mechanical energy conversion for power in terms of voltage for a two-phase synchronous reluctance motor:

$$P_{reluctance} = \left(\frac{1}{x_q} - \frac{1}{x_d}\right)\frac{V^2}{2}sin2\delta. \qquad \text{(D-14)}$$

The demonstration is as follows. According to Blondel theorem in steady state operation of a synchronous machine the voltage equation for each phase is:

$$E = V + rI + ix_d I_d + ix_q I_q \ , \qquad \text{(D-15)}$$

where $E$ is the induced voltage (rms value) in each phase of the stator depending on the field current, $V$ is the applied voltage (rms value) from an alternating current source, $x_d$ and $x_q$ are the rectances in axis d and q respectively, $i = \sqrt{-1}$ , $I_q$ and $I_d$ are components of the phase current $I$ as shown in phasor diagram of Fig. D2.

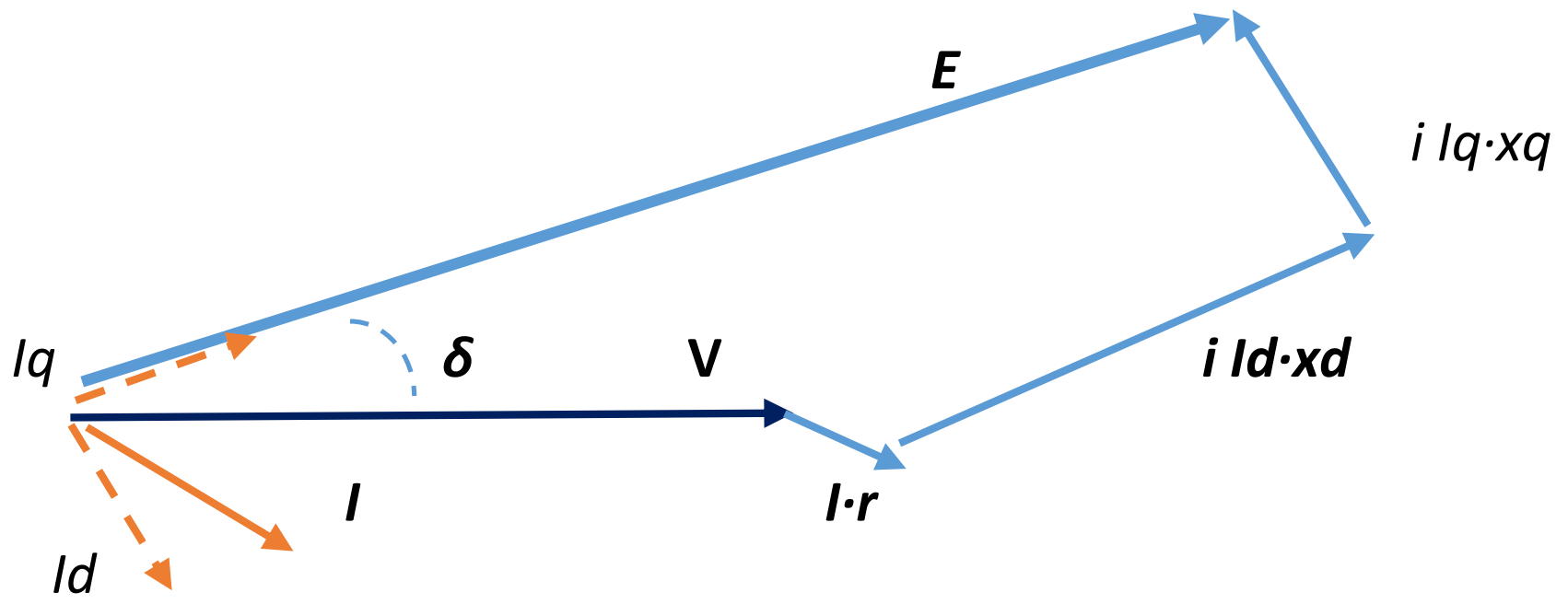

Figure D2. Phasor diagram.

From this phasor diagram if the resistance r may be neglected:

$$I_d = \frac{E - V\cos\delta}{x_d} \qquad \text{(D-16)}$$

$$I_q = \frac{V\sin\delta}{x_q} \qquad \text{(D-17)}$$

In a reluctance synchronous motor there is no field current so that E=0. Therefore starting from the expression $P_M = \omega(L_q - L_d)I_d I_q = (x_q - x_d)I_d I_q$, we have:

$$P_M = -(x_q - x_d)\frac{V^2 \sin\delta\cos\delta}{x_d x_q} = -(x_q - x_d)\frac{V^2 \sin 2\delta}{2x_d x_q} \qquad \text{(D-18)}$$

Without considering the minus sign which depends on the convention for generator and motor, there is concordance between expressions (D-14) and (D-18) for the mechanical power developed by a reluctance synchronous motor.